\documentclass[]{pasj01}
\draft

\begin{document} 
\Received{}
\Accepted{}

\title{Radial-velocity search and statistical studies for short-period planets in the Pleiades open cluster}

\author{Takuya \textsc{Takarada},\altaffilmark{1, 2, 3}$^{*}$ 
Bun'ei \textsc{Sato},\altaffilmark{4}
Masashi \textsc{Omiya},\altaffilmark{1,}\altaffilmark{2}
Yasunori \textsc{Hori},\altaffilmark{1,}\altaffilmark{2}
Michiko S. \textsc{Fujii}\altaffilmark{5}
}
\altaffiltext{1}{Astrobiology Center, 2-21-1 Osawa, Mitaka, Tokyo 181-8588, Japan}
\altaffiltext{2}{National Astronomical Observatory of Japan, 2-21-1 Osawa, Mitaka, Tokyo 181-8588, Japan}
\altaffiltext{3}{Graduate School of Science and Engineering, Saitama University, Simo-okubo 255, Sakura-ku, Saitama 338-8570, Japan}
\altaffiltext{4}{Department of Earth and Planetary Sciences, School of Science, Tokyo Institute of Technology, 2-12-1 Ookayama, Meguro-ku Tokyo 152-8551}
\altaffiltext{5}{Department of Astronomy, Graduate School of Science, The University of Tokyo, 7-3-1 Hongo, Bunkyo-ku, Tokyo 1130033, Japan}
\email{takuya.takarada@nao.ac.jp}



\maketitle

\begin{abstract}
We report radial-velocity search for short-period planets in the Pleiades open cluster.
We observed 30 Pleiades member stars at the Okayama Astrophysical Observatory (OAO) with High Dispersion Echelle Spectrograph (HIDES). 
To evaluate and mitigate the effects of stellar activity on radial-velocity measurements, we computed four activity indicators (FWHM, $V_{\rm span}$, $W_{\rm span}$ and $S_{\rm H{\alpha}}$).
Among our sample, no short-period planet candidates were detected.
Stellar intrinsic RV jitter was estimated to be ${\rm 52\ m\,s^{-1}}$, ${\rm 128\ m\,s^{-1}}$ and ${\rm 173\ m\,s^{-1}}$ for stars with $v\sin i$ of ${\rm 10\ km\,s^{-1}}$, ${\rm 15\ km\,s^{-1}}$ and ${\rm 20\ km\,s^{-1}}$, respectively.
We determined the planet occurrence rate from our survey and set the upper limit to 11.4\% for the planets with masses 1--13 $M_{\rm JUP}$ and period 1--10 days.
To set a more stringent constraint on the planet occurrence rate, we combined the result of our survey with those of other surveys targeting open clusters with ages between 30--300 Myr. 
As a result, the planet occurrence rate in young open clusters was found to be less than 7.4\%, 2.9\% and 1.9\% for the planets with an orbital period of three days and masses between 1--5, 5--13, and 13--80 $M_{\rm JUP}$, respectively. 
\end{abstract}

\section{INTRODUCTION}
Over 4,000 planets have been discovered since the discovery of the first known exoplanet, 51 Pegasi b \citep{Mayor1995}, and the increasing number of detections helps us to understand the statistical properties of planets and develop theoretical models of planetary formation and evolution. 
One of the clustered groups among them is the hot Jupiters (HJs), giant planets with masses greater than 0.3 $M_{\rm JUP}$ and orbital periods of $\simeq1-10$ d.
HJs have peculiar orbital properties compared to other planets in the solar system, and thus many researchers have tried to unravel the origins of HJs. 
However, the formation and evolution of HJs are still not completely understood.
In the standard core-accretion scenario, giant planets are believed to be formed beyond the snow line, far from the central star, where there exists plenty of solid materials to form the massive solid cores of giant planets (e.g, \cite{Pollack1996}). 
The existence of HJs suggest there should be some mechanisms by which the giant planets are displaced from their birthplace to the vicinity of their central star. 

Two leading models are proposed to explain the formation of HJs -- disk-driven migration (DDM, so-called Type-II migration) and high-eccentricity tidal migration (HEM). 
In DDM, the young giant planet creates a gap in the protoplanetary disk and moves inward through the gap (e.g. \cite{Goldreich1980}). 
In HEM, giant planets are scattered through gravitational interaction with other companions on high-eccentricity orbits that evolve into close-in circular orbits through a tidal interaction with the central star (e.g., \cite{Rasio1996}).
The crucial factor to discriminate between DDM and HEM is the time-scale over which HJs are formed by the respective mechanisms. 
In DDM, the planet migration has to be completed before the dissipation of the protoplanetary disk, and the timescale of HJ formation is determined by the lifetime of the disk ($\lesssim10$ Myr, e.g., \cite{Haisch2001}). 
In HEM, planet scattering and tidal circularization take hundreds of Myr to form the final shape of the system. 
Therefore, we can consider that the planet survey targeting stars in the age groups of approximately 100 Myr will help to distinguish between DDM and HEM. 
In this respect, open clusters are optimal targets, since stars in open clusters can be assumed to be born at the same time and location, and with identical chemical compositions.
This allows us to precisely determine stellar properties such as age, mass, and metallicities. 
Furthermore, stars in open clusters are typically younger ($\lesssim1$ Gyr) than the field stars that were the subject of interest in past planet surveys.

Several research groups have performed planet surveys in open clusters using the RV method. 
In the Hyades open cluster, an eccentric HJ, HD 285507 b, was reported by \citet{Quinn2014}.
A distant giant planet was also found orbiting an evolved star, $\epsilon$ Tau, which was the first planet detected in an open cluster \citep{Sato2007}. 
In the Praesepe open cluster, two HJs, Pr0201 b and Pr0211 b \citep{Quinn2012}, and one distant companion, Pr0201 c \citep{Malavolta2016}, were discovered.
In M 67, three HJs and two distant planets were discovered (\cite{Brucalassi2014}, \yearcite{Brucalassi2016}, and \yearcite{Brucalassi2017}).
Recently, \citet{Delgado2018} reports one distant companion in IC 4651.
The first transiting planets in open clusters were found in NGC 6811 with the Kepler space telescope \citep{Meibom2013}. 
Since starting the K2 mission, several smaller transiting planets have been discovered in Hyades (e.g., \cite{Mann2016}), Praesepe (e.g., \cite{Mann2017}), and Ruprecht 147 \citep{Curtis2018}.
Although several HJs have been discovered in open clusters, the ages of these clusters ($\gtrsim$600 Myr) are relatively older than or comparable to the tidal circularization timescale, and the constraint on the formation and evolution mechanism of HJs has been inconclusive. 
As for the younger open clusters, \citet{Paulson2006} and \citet{Bailey2018} have used the RV method to conduct planet surveys targeting younger clusters. 
\citet{Paulson2006} observed 33 stars in several open clusters in the age groups of 12--300 Myr, and \citet{Bailey2018} observed 138 stars in NGC 2516 (141 Myr) and NGC 2422 (73 Myr). 
In the K2 mission, some young open clusters (Pleiades, M 18, M 21, M 25, and M 35) in age groups less than 200 Myr were observed. 
Up until today, no planets have been discovered in these observations.

In this paper, we report a short-period planets survey in the Pleiades open cluster ($\sim100$ Myr) using RV methods. 
The young age of the cluster enables us to set more stringent constraints on the timescale of formation of HJs. 
The outline of this paper is as follows. 
In section 2, we describe our stellar samples and observations. 
The analysis and its results are given in section 3 and 4, respectively. 
Then, we discuss stellar RV jitter and short-period planet occurrence rates in young open clusters in section 5.
Finally, section 6 presents summary of this paper.
\section{SAMPLE AND OBSERVATION}
We selected stars from the Pleiades open cluster.
After Hyades, the Pleiades open cluster is the second nearest star cluster ($133.5\pm1.2$ pc, \cite{Soderblom2005}; $120.2\pm1.9$ pc, \cite{Leeuwen2009}) to our solar system. 
Because of its proximity and population ($\sim$1300 stars, \cite{Lodieu2012}), the Pleiades has been a subject of several studies. 
The age of the cluster was determined to be 115$\pm$5 Myr by the lithium depletion boundary technique \citep{Dahm2015}. 
Although different studies gave different ages for the Pleiades (e.g., 125$\pm$5 Myr, \cite{Stauffer1998}), all the ages were around 100 Myr. 
As for the metallicity, \citet{Funayama2009} spectroscopically derived [Fe/H]$=$0.03$\pm$0.05 dex from the Fe\,\emissiontype{I} and Fe\,\emissiontype{II} lines of 22 A-, F-, and G-type stars of Pleiades. 
\citet{Gebran2008} also derived [Fe/H]$=$0.06$\pm$0.02 dex from the Fe\,\emissiontype{II} lines of five F-type stars in the cluster. 
Thus, we can consider the metallicities of the Pleiades member stars as being close to the solar value. 
The fundamental parameters of the Pleiades are listed in table \ref{tab_plei}.
Planet searches in Pleiades have been performed by transit and direct imaging methods. 
In the K2 mission, 1,000 candidate Pleiades member-stars (411 FGK dwarfs and 603 M dwarfs) were observed in the Campaign field 4 (e.g., \cite{Gaidos2017}), but no planets were detected. 
\citet{Geissler2012} and \citet{Rodriguez2012} using the direct imaging technique detected distant brown dwarf companions around the HII 1348 and HD 23514, respectively. 
So far, no planets have been detected around the Pleiades member-stars.
\begin{table}
\tbl{Fundamental parameters of the Pleiades open cluster}{%
 \begin{tabular}{lcr}
 \hline
Parameter&Value&Reference\\\hline
Distance (pc) 		&	$133.5\pm1.2$ &	\cite{Soderblom2005}\\
				&	$120.2\pm1.9$	& 	\cite{Leeuwen2009}\\\hline
Age (Myr)			&	125$\pm$5	&	\cite{Stauffer1998}	\\
	 			&	115$\pm$5	&	\cite{Dahm2015}	\\\hline
${\rm [Fe/H]}$ (dex)	&	0.06$\pm$0.02	&	\cite{Gebran2008}		\\
				&	0.03$\pm$0.05	&	\cite{Funayama2009}	\\\hline
Member			&	$\sim$1300	&	\cite{Lodieu2012}\\\hline
 \end{tabular}}
\label{tab_plei}
\end{table}

Stars were selected based on the following criteria: i) the membership probability is over 70\% (\cite{Kharchenko2004}; \cite{Bouy2015}); ii) visual (V) magnitude is approximately 9.43--10.6; and, iii) the spectral type is between F8 and G5. 
The faint end of the V magnitude is determined by the feasibility of RV measurements with a small aperture telescope. 
The third criterion is considered not to include any early-type stars that typically rotate at high velocities. 
On these criteria, we observed 30 stars. 
%
%
%
%

All data were obtained with the 1.88-m reflector and High Dispersion Echelle Spectrograph (HIDES; \cite{Izumiura1999}, \cite{Kambe2013}) at the Okayama Astrophysical Observatory (OAO). 
An iodine absorption cell (I$_{\rm 2}$ cell; \cite{Kambe2002}) was set on the optical path, which superposes numerous iodine absorption lines onto a stellar spectrum of $5000$--$5800\,{\rm \AA}$ as the fiducial wavelength reference for precise RV measurements. 
After the 2007 December upgrade, HIDES has three CCDs from one CCD; the wavelength coverage of each CCD is $3700$--$5000\,{\rm \AA}$ (CCD-blue), $5000$--$6300\,{\rm \AA}$ (CCD-green), and $6300$--$7500\,{\rm \AA}$ (CCD-red). 
This enables us to evaluate stellar activity levels (i.e. H$\alpha$ line) with CCD--red and line profile variations with CCD-blue simultaneously with stellar RV variations measured by using CCD-green. 
We obtained stellar spectra with the high-efficiency fiber-link mode of HIDES. 
In the fiber-link mode, the width of the sliced image was {1.05''}, which corresponds to the spectral resolution $(R=\lambda/\Delta \lambda)$ of 55,000 by about 3.8-pixels sampling. 
With this setting, we can typically obtain a signal-to-noise ratio (SNR) of $\sim$60 pixel$^{-1}$ at 5500 \AA\ with an exposure time of 30 minutes for 10.0 mag stars.

The reduction of echelle data was performed with IRAF\footnote{IRAF is distributed by the National Optical Astronomy Observatories, which is operated by the Association of Universities for Research in Astronomy, Inc. under cooperative agreement with the National Science Foundation, USA.} in the conventional way.
For CCD-blue, the orders in the bluer region (shorter than $4400\,{\rm \AA}$) severely overlap with adjacent orders due to using image slicer, and thus we cannot subtract scattered light in the region.
Therefore we divided spectrum into a red part ($4400$--$5000\,{\rm \AA}$ without order overlapping) and blue part ($3700$--$4400\,{\rm \AA}$ with order overlapping).
In the red part, we subtract scattered light with IRAF task, apscatter, and performed line-profile analysis.
\section{ANALYSIS}
\subsection{Radial velocity and rotational velocity}
\label{sec:rv}
For precise RV measurements, we use spectra covering $5000$--$5800\,{\rm \AA}$ on which I$_{\rm 2}$ lines are superimposed by I$_{\rm 2}$ cell.
We compute RV variations, following the method described in \citet{Sato2002}, which is based on the ones established by \citet{Valenti1995} and \citet{Butler1996}.
In this method, the observed spectrum (I$_2$-superposed stellar spectrum) is modeled with the use of both a stellar template spectrum and high resolution I$_2$ spectrum which are convolved with instrumental profile (IP) of the spectrograph.
IP is modeled as a sum of one central gaussian and ten satellite gaussians whose widths and positions of the gaussians are fixed and their heights are treated as free parameters.
A 2nd-order polynomial is used for the wavelength scaling.
The observed spectrum is divided into hundreds of segments with typical width of 150$\>$pixel.
The final RV value is derived by taking the average of RV values of all the segments.

In our observing runs, we could not obtain pure stellar spectra used for templates  because of their faintness and limited observation time.
Therefore, we used the spectra obtained by other instruments.
Among our sample, ten and five stars were observed through High Accuracy Radial Velocity Planet Searcher (HARPS) with the 3.6 meter telescope at La Silla and High Resolution Echelle Spectrometer (HIRES) with the 10 meter telescope at Keck Observatory, respectively.
Additionally, we obtained the spectra of eight stars through High Dispersion Spectrograph (HDS) with the 8.2 meter telescope at Subaru Telescope.
Excluding the duplication, we collected template spectra for 15 stars.
Since the wavelength resolutions of each spectrograph are sufficiently high compared to the typical rotational velocities of our sample, we did not perform deconvolution for these spectra. 
For stars without pure spectra, we adopted the template spectrum of a sample with a similar rotational velocity.

To determine a template spectrum for stars without pure spectra, we derive the rotational velocity of our sample.
We fit the observed spectra covering $6400$--$6500$\,\AA\ with theoretical spectra made by SPECTRUM code \citep{Gray1994}.
The chemical composition of the model spectra is assumed to be identical to that of the Sun.
The fitting parameters are the rotational velocity, $v\sin i$, Doppler shift, and a linear normalization function.
The parameter step of $v\sin i$ is fixed at $5\ {\rm km\,s^{-1}}$.
We adopt the linear limb-darkening coefficient and fix it at 0.6.
Following \citet{Gray2005}, the modeled spectrum can be written as 
\begin{equation}
I(v) = k[M(v)\ast S(v+\Delta v)],
\end{equation}
where $k$, $\Delta v$, and $M(v)$ are a linear normalization function, Doppler shift, and disk-integrated broadening function due to stellar rotation, respectively.
$M(v)$ is defined as
\begin{equation}
M(v)=\frac{2(1-\epsilon)[1-(v/v_{\rm Rot})^2]^{1/2}+\frac{1}{2}\pi\epsilon[1-(v/v_{\rm Rot})^2]}{\pi v_{\rm Rot}(1-\epsilon/3)}
\end{equation}
where $\epsilon$ is the linear limb-darkening coefficient, and $v_{\rm Rot}$ is the maximum rotational velocity in the stellar disk and is equivalent to $v\sin i$.
\subsection{Stellar activity indices}
\label{sec:BIS}
When the RV variations are caused by stellar activities (e.g., cool spots), the line profiles are deformed at the same time as the occurrence of the RV variations.
Therefore, we performed a line-profile analysis to evaluate the effect of stellar activities on RV variations.
In order to enhance the SNR of line profile, we derive mean stellar absorption line profiles with the use of the least-squares deconvolution (LSD) method following \citet{Kochukhov2010}.
To derive LSD profile, we refer to the VALD line-list (http://vald.astro.uu.se/$\sim$vald/php/vald.php).
Then, we computed three line-profile indices from a mean line profile; the FWHM, $V_{\rm span}$, and $W_{\rm span}$.
The definition of each indices were referred to \citet{Santerne2015}.

\begin{figure}
\begin{center}
\includegraphics[width=12cm]{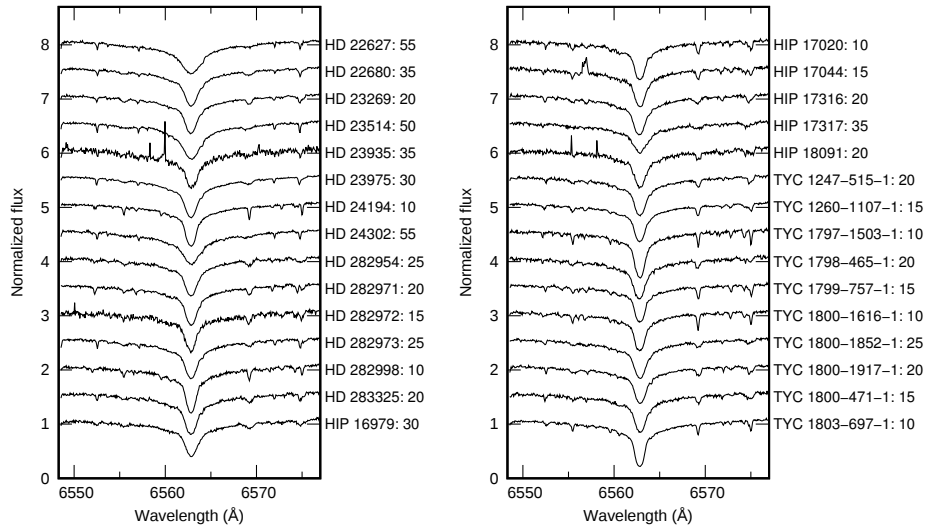}
\end{center}
\caption{H$\alpha$ lines of our sample. Stellar name and $v\sin i$ in ${\rm km\ s^{-1}}$ are shown on the right-hand side of each spectrum. For clarity, a vertical offset is added to each spectrum.}
\label{halpha1}
\end{figure}

\begin{figure}
\begin{center}
\includegraphics[width=10cm]{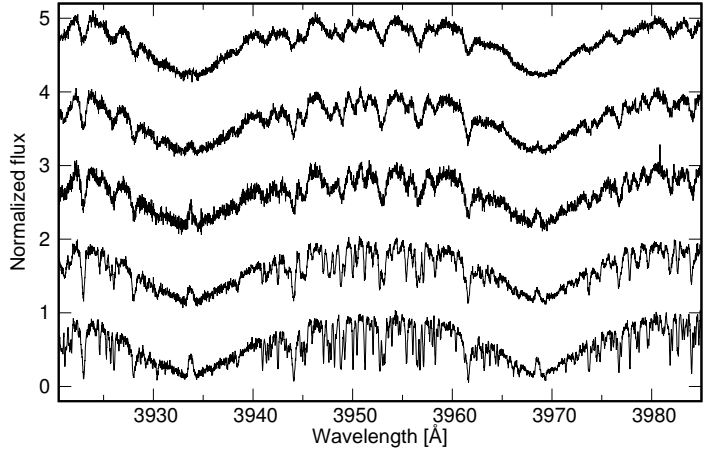}
\caption{Ca\,\emissiontype{II} H (at 3968.47\,\AA) and K (at 3933.66\,\AA) lines for our sample. From top to bottom: HD 24302 ($v\sin i=55\ {\rm km\,s}^{-1}$), HD 22680 ($35\ {\rm km\,s}^{-1}$), HIP 16979 ($30\ {\rm km\,s}^{-1}$), HIP 17044 ($15\ {\rm km\,s}^{-1}$), and TYC 1800-1616-1 ($10\ {\rm km\,s}^{-1}$). For clarity, a vertical offset of 1.0 is added to each spectrum. The spectra are retrieved from PHASE3 ARCHIVE INTERFACES (http://archive.eso.org/wdb/wdb/adp/phase3\_spectral/form).}
\label{caharps}
\end{center}
\end{figure}

We additionally check against stellar activities with H$\alpha$ line.
To evaluate the flux variabilities of the H${\alpha}$ line core at 6562.808 \AA, we define the H${\alpha}$ index, $S_{\rm{H{\alpha}}}$, as
\begin{equation}
S_{\rm{H{\alpha}}}=\frac{F_{\rm{H{\alpha}}}}{F_{\rm{B}}+F_{\rm{R}}},
\end{equation}
where $F_{\rm H{\alpha}}$ is the integrated flux within a 1.6-\AA-wide bin centered at the H${\alpha}$ lines, and $F_{\rm B}$, $F_{\rm R}$ is a 6.0-\AA-wide bin centered at 6552.3 \AA\ and 6574.8 \AA, respectively.
The band width of $F_{\rm H{\alpha}}$ follows the definition of \citet{Gomes2011}, and the band width and center of $F_{\rm B}$ and $F_{\rm R}$ are newly defined since the wavelength coverage differs from the one used in \citet{Gomes2011}.
In figure \ref{halpha1}, we show the spectra, including H$\alpha$ lines, of our sample.
For reference, the template spectra covering Ca\,\emissiontype{II} H and K lines are shown in figure \ref{caharps}. 
We can see that there are significant reversal emissions in the line cores. This indicates that our sample are chromospherically active.

\subsection{Statistical analysis}
Based on the RV values and activity indices, we determine a planet occurrence rate.
First, we derive the detection limit with the root-mean-square (RMS) method \citep{Meunier2012}.
The procedure of the RMS method is described below:
\begin{enumerate}
\item For each of our sample star, we assume the existence of a planet with a given minimum mass and orbital period.
\item We generate RV signals caused by the assumed planets at each observational epoch. In this step, the orbital phase is changed randomly. 
\item We compute the RMS of the simulated RV variations. If the RMS of the simulated RVs is greater than the RMS of the observed RVs and median of the RV measurement errors, the assumed planet is considered to be detected.
\item We repeat steps 1--3 1,000 times. When all the assumed planets are detected, the given pair of planet mass and orbital period are adopted as a detection limit.
\end{enumerate}

Next, we compute the search completeness, $C$, by stacking the detection limits of our sample \citep{Howard2010}.
We define $C$ as
\begin{equation}
C=\frac{1}{N}\sum_{i=1}^{N}\delta_i
\end{equation}
where $N$ is the number of stars and $\delta_i$ represents whether the assumed planet is detectable; we set $\delta_i=1$ if the assumed planet mass at a given orbital period is above the detection limit, and $\delta_i=0$ if not.
The search completeness shows the fraction of stars from which we can detect planets with a given mass and orbital period if present.

Finally, we compute the planet occurrence rate following \citet{Borgniet2017}.
For the first step, we compute the integrated search completeness, $C_I$, over a specific parameter range, $[m_{p1}:m_{p2}]$ and $[P_1:P_2]$, defined as
\begin{equation}
\label{stat1}
C_I=\frac{\displaystyle \sum_{P_1}^{P_2}\sum_{m_{p1}}^{m_{p2}}\ \left(\frac{1}{N}\sum_{i=1}^{N}\delta_i\right)\ {\rm d}P~{\rm d}m_p}{\displaystyle \sum_{P_1}^{P_2}\sum_{m_{p1}}^{m_{p2}}\ {\rm d}P~{\rm d}m_p}.
\end{equation}
In many cases, the search completeness is not 100\%.
To account for the search incompleteness, we estimate the number of missed planets, $n_{\rm miss}$, defined as
\begin{equation}
\label{stat2}
n_{\rm miss}=n_{\rm det}\times\left(\frac{1}{C_I}-1\right)
\end{equation}
where $n_{\rm det}$ is the number of detected planets within the specific parameter range.
Although we did not detect any planets in our survey, we defined $n_{\rm det}=1$ following \citet{Borgniet2017} to set an upper limit.
Now we can compute the planet occurrence rate with the use of binomial statistics.
The probability $f$ of drawing $n_{\rm det}$ systems with at least one planet from $N$ sample stars can be written as
\begin{equation}
\label{stat3}
f(n_{\rm det},N,p)=
\left(
  \begin{array}{c}
     N \\
     n_{\rm det}
  \end{array}
 \right)
p^{n_{\rm det}}(1-p)^{N-n_{\rm det}}
\end{equation}
where $p$ is the probability of having at least one planet around one star and is equivalent to the planet occurrence rate.
The most reasonable planet occurrence rate is given by the value at which the probability, $f$, gets highest.
The 1$\sigma$ and 2$\sigma$ uncertainties are given at 68.27\% and 95.45\% of $f$, respectively.
Then, we correct the search incompleteness by multiplying the derived $p$ with ${(n_{\rm det}+n_{\rm miss})}/{n_{\rm det}}$.
\section{RESULTS}
\subsection{Radial-velocity variations}
\begin{figure}[htb]
\begin{center}
\includegraphics[width=0.7\textwidth]{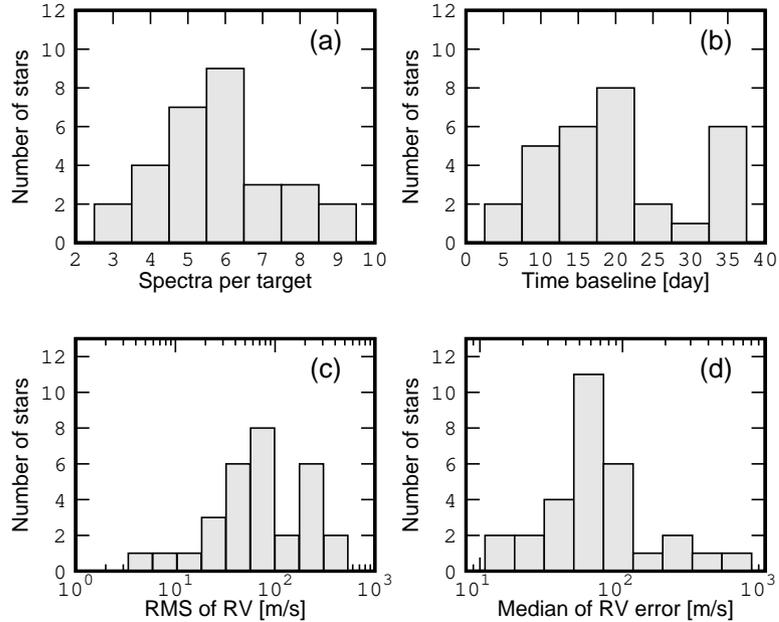}
\end{center}
\vspace{-0.7cm}
\caption{Properties of our observation. (a), (b), (c) and (d) correspond to the distributions of spectrum number per target, observation time baseline, RMS of RV, and median of RV measurement error, respectively.}
\label{rvstat0}
\end{figure}

In our survey, we planned to observe at least 5 times for each star: three consecutive nights and two consecutive nights with an interval of several days.
With the limited observing time, this is an optimal data sampling to establish the presence of HJs with orbital periods of 1--10 days.
As shown in figure \ref{rvstat0} (a) and (b), we made observations in this way for most of the samples.
Even for the stars with fewer observations, we succeeded in conducting a continuous three-days observation.
Figure \ref{rvstat0} (c) and (d) show the histograms of the RV computation results.
RMS of the measured RVs and median of RV measurement errors are shown in table \ref{restab}.

\begin{figure}
\begin{center}
\includegraphics[width=0.7\textwidth]{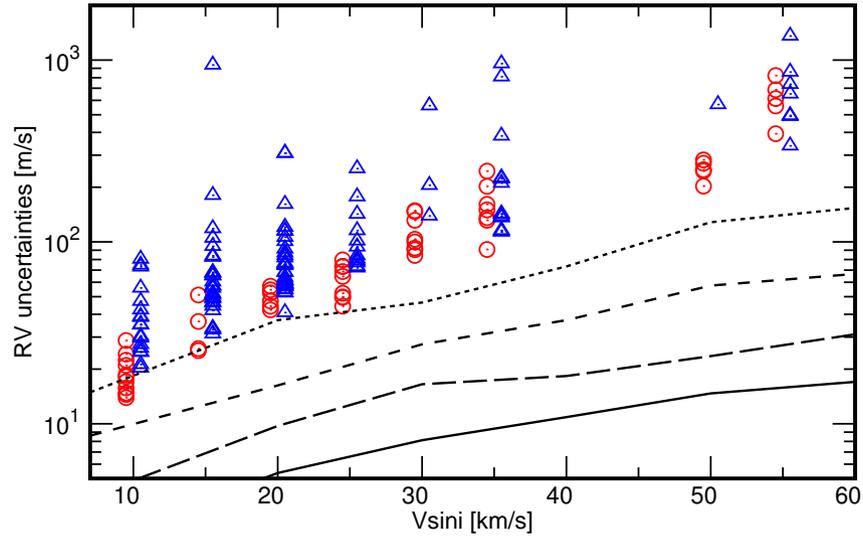}
\end{center}
\caption{RV measurement errors as a function of stellar rotational velocities. Red circles and blue triangles are the results for data with SNR $>$50 and $<$50, respectively. From top to bottom, each line shows the expected RV measurement errors for SNR with 30, 50, 100 and 200, respectively. The expected errors are determined by use of model spectra to which photon noise is added based on an assumed SNR.}
\label{rvstat1}
\end{figure}

\begin{figure}
\begin{center}
\includegraphics[width=0.8\textwidth]{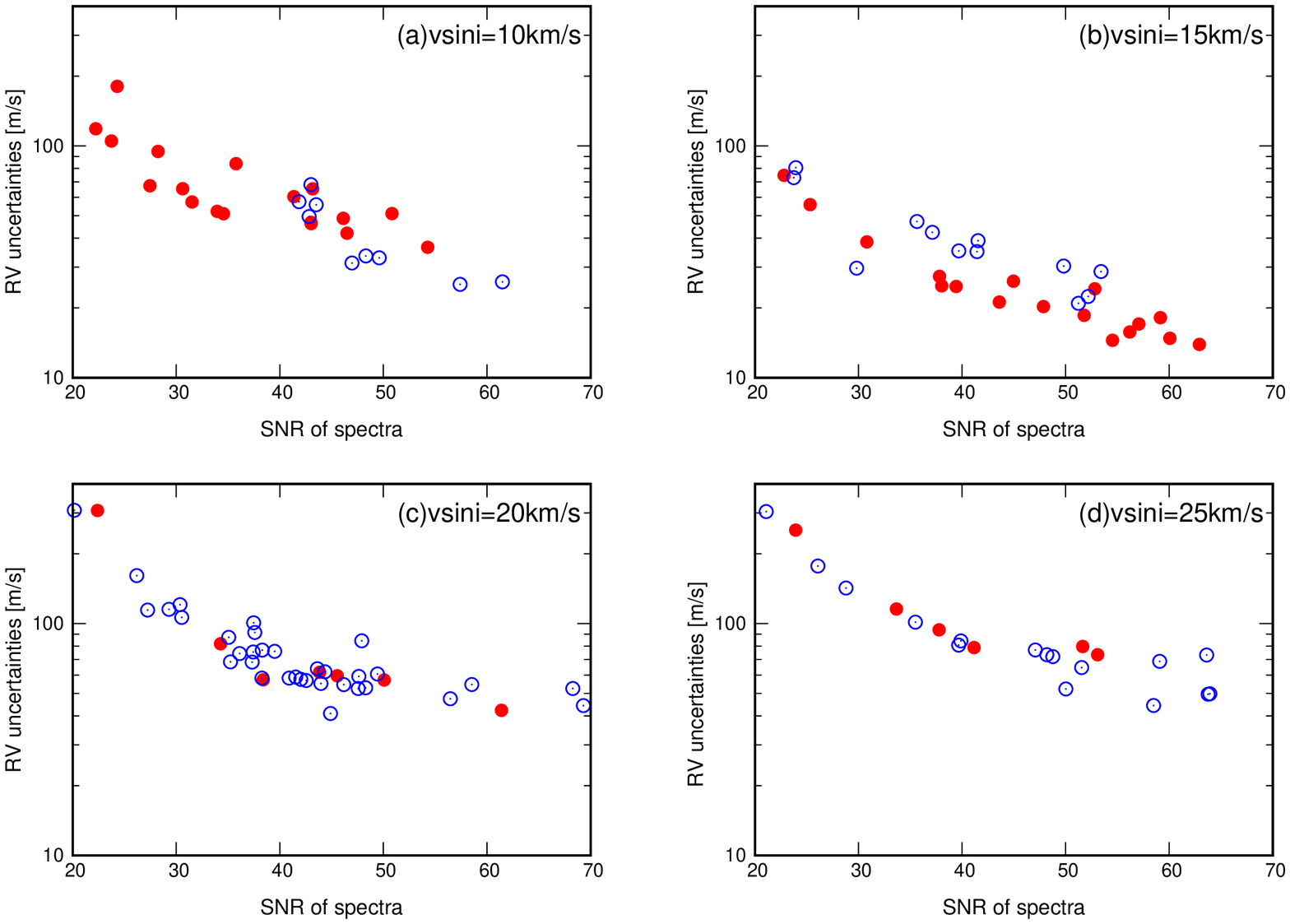}
\end{center}
\caption{RV uncertainties as a function of SNR of observed spectra at $\sim$5500 \AA. Red and blue circles are the values derived with the stellar templates of the stars themselves and those of other stars, respectively. (a), (b), (c) and (d) are the result for stars with $v\sin i=10,\ 15,\ 20,\ {\rm and}\ 25\ {\rm km\,s^{-1}}$, respectively. }
\label{rvstat2}
\end{figure}

Figure \ref{rvstat1} shows the RV measurement errors ($\sigma_{\rm RVerr}$) derived from our observations compared with the expected $\sigma_{\rm RVerr}$.
Here, the expected $\sigma_{\rm RVerr}$ is derived by performing RV measurements for mock spectra to which photon noise is added base on the assumed SNR.
We see that $\sigma_{\rm RVerr}$ exceeds $100\ {\rm m\,s^{-1}}$ for stars with $v\sin i\geq30\ {\rm km\,s^{-1}}$.
When compared with the expected $\sigma_{\rm RVerr}$, the observed $\sigma_{\rm RVerr}$ seems to be worse by a factor of 2--5.
This indicates that there might be some factors that led to the gap between the RV measurement errors and the expected errors.
Two possible factors can be considered as inducing the discrepancy of RV errors.
Firstly, the stellar template mismatch could be a matter of RV precision for stars for which we used template spectra of different stars.
Figure \ref{rvstat2} shows $\sigma_{\rm RVerr}$ derived with the use of the templates of the star itself and those of other stars.
From the figure \ref{rvstat2}, no significant discrepancies are seen.
This implies that the use of different stellar spectra as templates causes no serious problems.
Secondly, the wavelength-dependent stellar activities may also worsen the RV precision.
\citet{Feng2017} presented the evidence for wavelength-dependent RV noise for the inactive star, $\tau$ Ceti, and this should also be the case for young active stars.
If each stellar absorption line profile was subjected to different deformations, the velocity variance of each segment would increase.

\begin{figure}
\begin{center}
\includegraphics[width=10cm]{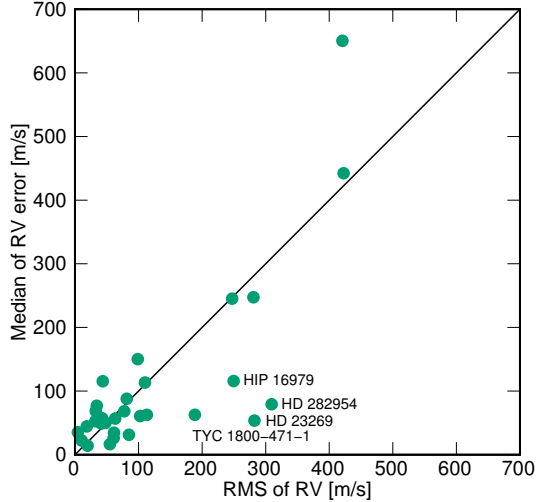}
\end{center}
\caption{Median of RV measurement errors as a function of RMS of RVs. The solid line represents an one-to-one relation.}
\label{spec_cont3}
\end{figure}

Figure \ref{spec_cont3} illustrates the comparison between the median of RV measurement errors and the RMS of RVs.
We can see that four stars (TYC 1800-471-1, HD 23269, HIP 16979, and HD 282954) exhibit large RV variations when compared to the RV measurement errors.
The large RV variations could be caused by planets or stellar activities.
Hence RVs of these stars are compared with activity indices, from which we found significant correlations between them.
For these reasons, we conclude that the significant RV variations seen from TYC 1800-471-1, HD 23269, HIP 16979, and HD 282954 were caused by stellar activities and no planet candidates were detected among our sample.
Results of RV measurements and activity indices are listed in table \ref{restab} and shown in appendix.

\begin{table}
\tbl{Results of RV measurements and activity indices.\footnotemark[$*$]}{%
\tabcolsep3pt 
 \begin{tabular}{lcccccccccccccc}\hline
Star 	&Sp. type&$V$& $v\sin i$	 &$\sigma_{\rm RV}$&$\sigma_{\rm RV_{err}}$&$\sigma_{\rm FWHM}$&$\sigma_{S_{\rm{H}{\alpha}}}$&$\sigma_{V_{\rm span}}$&$\sigma_{W_{\rm span}}$&$\sigma_{\rm photo}$&$M_{\rm p,3d}$&$M_{\rm p,10d}$	&$N_{\rm obs}$& $T_{\rm span}$\\
 &&(mag)& (${\rm km\ s^{-1}}$)&(${\rm m\ s^{-1}}$)&(${\rm m\ s^{-1}}$)& (${\rm km\ s^{-1}}$)&($\times 10^3$) &(${\rm km\ s^{-1}}$)&(${\rm km\ s^{-1}}$)&($\times 10^3$)	&($M_{\rm JUP}$)&($M_{\rm JUP}$)	& & (day)\\\hline
HD 23514 &F5&9.43& 50 & 281 & 247 & 0.364 & 1.276 & 0.212 & 0.500 & 3.310 & 3.7 & 7.9 & 6 & 10 \\
HD 24302 &F8&9.48& 55 & 421 & 650 & 0.861 & 1.095 & 1.147 & 1.319 & 1.782 & 20.0& 20.0 & 6 & 16 \\
HD 23935 &F8&9.58& 35 & 99 & 150 & 0.399 & 0.562 & 0.408 & 0.500 & -- & 7.9 & 3.1 & 5 & 16 \\
HD 23975 &G0&9.73& 30 & 110 & 113 & 0.180 & 0.304 & 0.557 & 1.007 & 1.766 & 20.0 & 3.0 & 7 & 21 \\
HD 22627 &G0&9.82& 55 & 423 & 442 & 0.990 & 0.583 & 0.553 & 1.516 & -- & 5.7 & 8.8 & 6 & 10 \\
HIP 17316 &G0&9.84& 20 & 41 & 49 & 0.198 & 0.697 & 0.181 & 0.422 & 3.533 & 0.7 & 1.1 & 8 & 19 \\
HIP 16753 &--&9.86& 25 & 48 & 50 & 0.072 & 0.639 & 0.067 & 0.597 & 1.997 & 3.4 & 2.0 & 5 & 19 \\
HD 282973 &G0&9.87& 25 & 34 & 77 & 0.304 & 1.761 & 0.251 & 0.534 & 2.372 & 0.9 & 2.6 & 7 & 16 \\
HD 22680 &G&9.93& 35 & 44 & 116 & 0.307 & 1.258 & 0.279 & 0.538 & -- & 1.6 & 2.1 & 8 & 24 \\
TYC 1260-1107-1 &--&9.99& 15 & 85 & 31 & 0.324 & 0.800 & 0.264 & 0.985 & 4.566 & 0.5 & 0.9 & 5 & 34 \\
TYC 1803-697-1 &G0&9.99& 10 & 55 & 16 & 0.185 & 0.424 & 0.014 & 0.196 & -- & 0.2 & 2.7 & 4 & 21 \\
HD 23269 &F8&10.0& 20 & 282 & 54 & 0.274 & 1.546 & 0.659 & 0.929 & 4.047 & 0.7 & 1.1 & 6 & 36 \\
HIP 16979 &F8&10.1& 30 & 250 & 116 & 0.476 & 1.408 & 0.650 & 0.590 & 7.350 & 1.6 & 2.2 & 6 & 34 \\
HD 24194 &G0&10.1& 10 & 20 & 14 & 0.307 & 0.281 & 0.251 & 0.437 & -- & 0.2 & 0.7 & 4 & 19 \\
TYC 1800-1852-1 &F9&10.1& 25 & 113 & 63 & 0.680 & 0.224 & 0.537 & 1.158 & 3.551 & 7.1 & 3.8 & 4 & 19 \\
TYC 1800-1917-1 &G0&10.2& 20 & 63 & 57 & 0.281 & 1.874 & 0.147 & 0.288 & 7.276 & 0.7 & 1.9 & 5 & 15 \\
HD 282971 &F8&10.2& 20 & 33 & 68 & 0.106 & 1.364 & 0.107 & 0.405 & 4.232 & 0.8 & 1.8 & 5 & 15 \\
TYC 1800-471-1 &F8&10.2& 15 & 189 & 63 & 0.408 & 1.942 & 0.217 & 0.656 & 4.204 & 1.7 & 4.1 & 6 & 20 \\
HD 282954 &--&10.3& 25 & 309 & 79 & 0.650 & 0.785 & 0.921 & 1.508 & 5.911 & 1.5 & 2.2 & 6 & 38 \\
TYC 1797-1503-1 &--&10.3& 10 & 11 & 22 & 0.030 & 1.519 & 0.047 & 0.567 & -- & 4.0 & 1.1 & 3 & 4 \\
HIP 17044 &--&10.3& 15 & 43 & 57 & 0.201 & 1.768 & 0.111 & 0.419 & -- & 1.1 & 1.7 & 9 & 27 \\
HIP 17317 &--&10.3& 35 & 247 & 245 & 1.312 & 1.013 & 0.951 & 2.080 & -- & 2.9 & 4.3 & 7 & 33 \\
HD 282972 &F9&10.4& 15 & 19 & 44 & 0.071 & 1.079 & 0.163 & 0.877 & 7.300 & 0.5 & 0.7 & 5 & 8 \\
HD 282998 &G0&10.4& 10 & 61 & 35 & 0.115 & 1.031 & 0.253 & 0.796 & 6.932 & 0.7 & 1.0 & 6 & 33 \\
TYC 1247-515-1 &F8&10.4& 20 & 102 & 61 & 0.437 & 0.504 & 0.137 & 0.345 & -- & 0.7 & 2.2 & 5 & 19 \\
HIP 18091 &--&10.5& 20 & 78 & 68 & 0.174 & 1.757 & 0.125 & 0.659 & -- & 0.9 & 1.4 & 8 & 31 \\
TYC 1799-757-1 &--&10.5& 15 & 32 & 53 & 0.225 & 0.774 & 0.115 & 0.291 & 6.552 & 0.6 & 3.2 & 4 & 11 \\
HIP 17020 &--&10.5& 10 & 5 & 35 & 0.148 & 1.390 & 0.093 & 0.581 & -- & 3.6 & 2.1 & 3 & 3 \\
TYC 1798-465-1 &--&10.6& 20 & 81 & 88 & 0.122 & 2.620 & 0.267 & 0.622 & -- & 1.2 & 1.6 & 6 & 8 \\
TYC 1800-1616-1 &G0&10.6& 10 & 61 & 26 & 0.441 & 1.067 & 0.491 & 0.674 & 4.959 & 0.7 & 1.0 & 9 & 34 \\\hline
 \end{tabular}}
 \begin{tabnote}
\hangindent6pt\noindent
 \hbox to6pt{\footnotemark[$*$]\hss}\unskip%
Star names, spectral type, V magnitude, $v\sin i$, RMS of RVs and median of RV measurement errors are given in columns 1--6, in that order. Columns 7--11 are the RMS of FWHM, $S_{\rm H{\alpha}}$, $V_{\rm span}$, $W_{\rm span}$ and relative flux, respectively. Columns 12--13 correspond to mass detection limit at 3 days and 10 days orbital period. Number of observations and time baseline are shown in the last two columns.
\end{tabnote}
\label{restab}
\end{table}

\subsection{Statistical analysis}
\begin{figure}
\begin{center}
\hspace{0.5cm}
\includegraphics[width=0.9\textwidth]{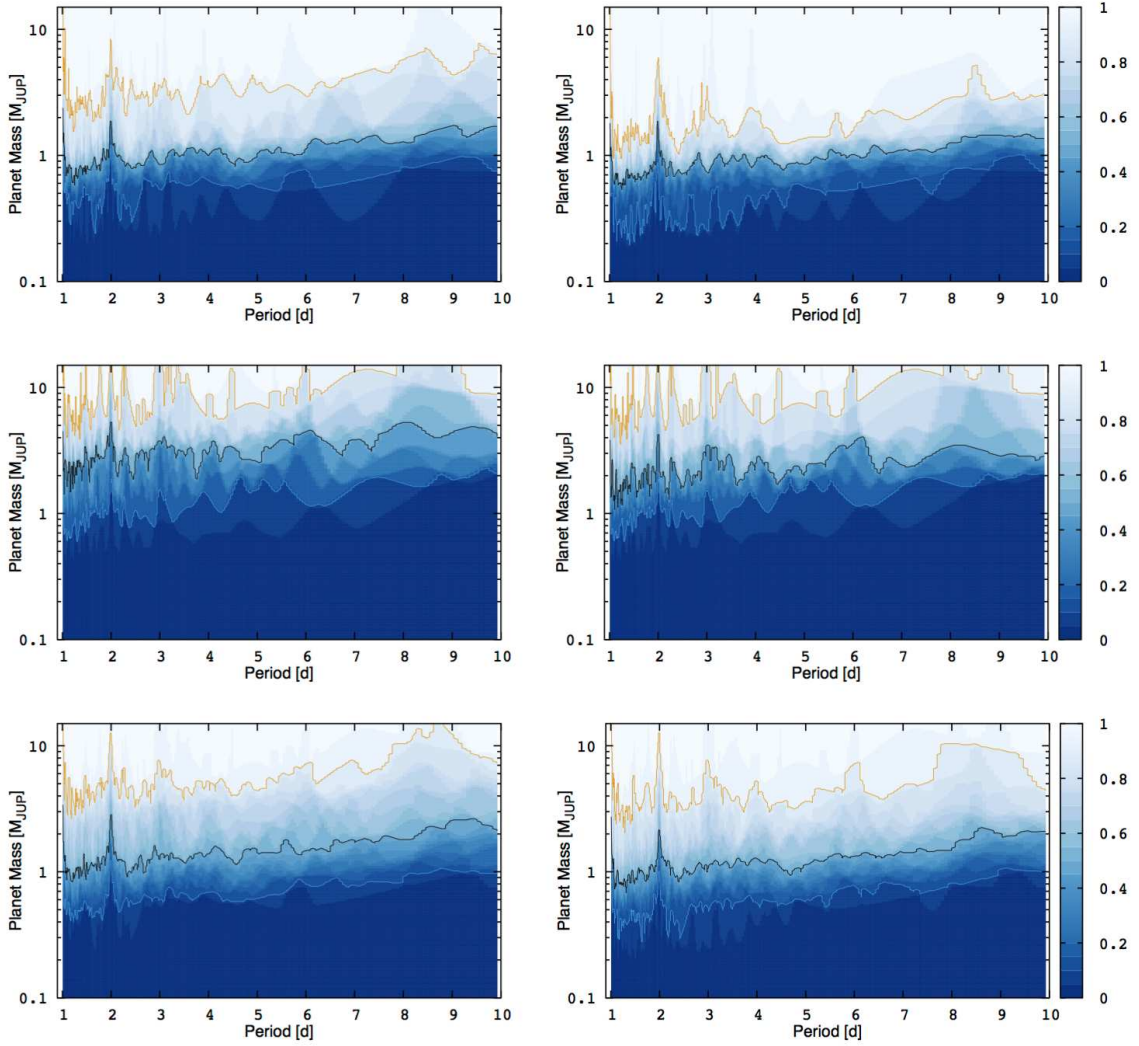}
\vspace{-0.5cm}
\end{center}
\caption{The completeness of our observation on a mass (x-axis)--orbital period (y-axis) plane. Color contour shows the fraction of stars satisfying detection limits well enough to rule out planets with a certain planet mass and orbital period. Orange, black and light blue lines correspond to 90\%, 50\% and 10\% completeness, respectively. Left column: the completeness of the slowly rotating sub-sample, the rapidly rotating sub-sample, and whole sample before adopting the correction of stellar activities (from top to bottom). Right column: same as the left one column but for those after adopting the correction.}
\label{detlim2}
\end{figure}

In order to determine the detection limits for our sample, the period range was divided into 250 points on logarithmic scale, from 0.5 to 11 days.
As for the planet mass range, we set 0.25 $M_{\rm JUP}$ as the minimum mass and increased the mass by 1\% up to 15 $M_{\rm JUP}$ as the maximum mass.
For the stars showing a significant correlation between the RVs and activity indicators, we fitted a linear function to the correlation and subtracted it to mitigate the effect of stellar activity on RV variations.
Then, the detection limits are computed again.
After that, we combined the detection limits of our sample to derive the search completeness of our survey (figure \ref{detlim2}).
Since stars with high rotational velocities show large RV measurement errors, we divided our sample into two sub-samples; 18 stars with $v\sin i<20\ {\rm km\,s}^{-1}$ and 12 stars with $v\sin i>20\ {\rm km\,s}^{-1}$ to examine the search completeness for less massive planets.
As for the sub-sample of slowly rotating stars (the top rows in figure \ref{detlim2}), the existence of planets with masses larger than 1 $M_{\rm JUP}$ was ruled out in half of the samples.
Furthermore, the correction of stellar activities improved the search completeness down to 3 $M_{\rm JUP}$ among 90\% of the samples.
We can see a remarkable improvement for the sub-sample of rapidly rotating stars (the middle rows in figure \ref{detlim2}).
Although the completeness is still worse than the sub-sample of slowly rotating stars, the correction lets us rule out massive close-in planets ($\sim10\ M_{\rm JUP}$) among 90\% of the samples.
From the whole sample (the bottom rows in figure \ref{detlim2}), we can see that our observation rules out the existence of massive HJs ($>5\ M_{\rm JUP}$) with orbital periods of 1--8 day.

Finally, we derived the planet occurrence rate from our survey.
The mass and period ranges are divided into two groups, respectively: 1--5 $M_{\rm JUP}$ and 5--13 $M_{\rm JUP}$ for mass, and 1--5 day and 5--10 day for orbital period.
Table \ref{tab_occ1} shows the occurrence rates thus derived.
The completeness in the domain of the less massive planets (1--5 $M_{\rm JUP}$) is lower compared to the domain of the massive planets.
This poor completeness value for the less massive planets is not only due to the small number of samples but also the low RV measurement precision, which prevents the detection of sub-Jupiter mass planets.
In the meantime, our survey distinctly investigates the domain of massive planets.
This indicates that RV-measurement precisions are sufficient to confidently rule out the existence of such planets among our sample.
However, errors in the occurrence rate are substantially large.
We expect that increasing the number of sample stars can make these errors smaller.

\begin{table}
\tbl{Planet occurrence rate in the Pleiades}{%
 \begin{tabular}{ccccccc}
 \hline
Mass & Period & Completeness & Missed planets & Occ. rate & 1$\sigma$ error & 2$\sigma$ error\\
$(M_{\rm JUP})$&(day)&(\%)& &(\%)&(\%) & (\%)\\\hline
1--5&1--5&79.5&0.25&4.0&12.5 & 17.6\\
1--5&5--10&68.9&0.45&4.6&14.4& 20.3\\\hline
5--13&1--5&95.7&0.04&3.3&10.4 & 14.6\\
5--13&5--10&92.0&0.09&3.5 & 10.8&15.2\\\hline
1--13&1--10&87.0&0.15&3.7&11.4&16.1\\\hline
 \end{tabular}}
\label{tab_occ1}
\end{table}
\section{DISCUSSION}
\subsection{Stellar radial-velocity jitter}
\label{sec_jitter}
Based on the analysis of RV and activity indicators, we evaluate the stellar RV jitter of the Pleiades stars.
The stellar RV jitter is defined as
\begin{equation}
\label{eq_jitter}
\sigma_{\rm jitter}=\sqrt{\sigma_{\rm RV}^2-\sigma_{\rm error}^2}
\end{equation}
where $\sigma_{\rm jitter}$ is the stellar RV jitter, $\sigma_{\rm RV}$ is the RMS of RV variations and $\sigma_{\rm error}$ is median of RV measurement errors.
We determined the stellar RV jitter at ${\rm 52\ m\,s^{-1}}$, ${\rm 128\ m\,s^{-1}}$ and ${\rm 173\ m\,s^{-1}}$ for stars with $v\sin i$ of ${\rm 10\ km\,s^{-1}}$, ${\rm 15\ km\,s^{-1}}$ and ${\rm 20\ km\,s^{-1}}$, respectively.
These values are consistent with those obtained by other surveys targeting open clusters as young as the Pleiades; ${\rm 128\ m\,s^{-1}}$ for NGC 2422 (73 Myr, \cite{Bailey2018}), ${\rm 67\ m\,s^{-1}}$ for NGC 2516 (141 Myr, \cite{Bailey2018}), ${\rm 58\ m\,s^{-1}}$ for Castor moving group (200$\pm$100 Myr, \cite{Paulson2006}) and ${\rm 65\ m\,s^{-1}}$ for Ursa Major moving group (300 Myr, \cite{Paulson2006}).

\begin{figure}
\begin{center}
\includegraphics[width=14cm]{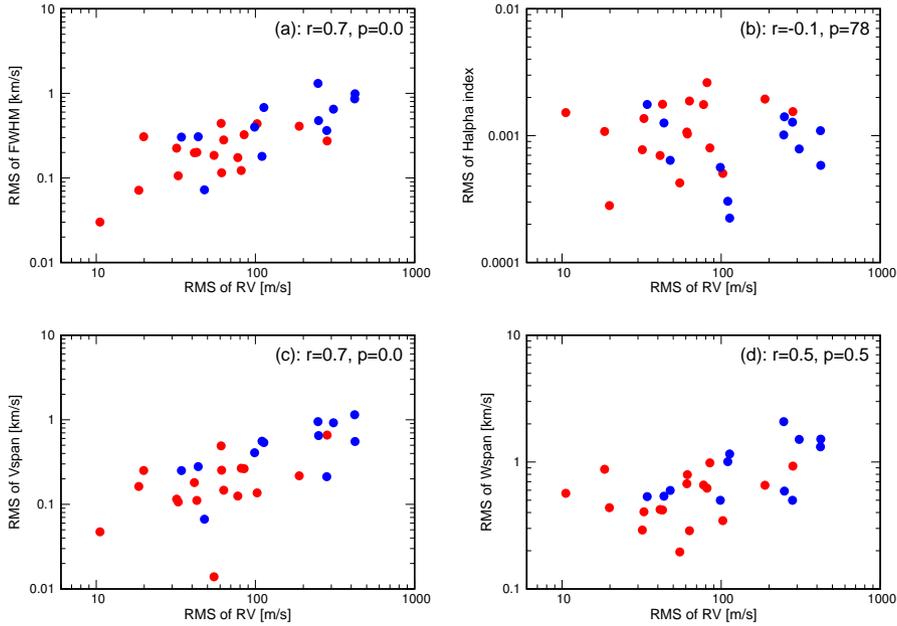}
\end{center}
\caption{RMS of activity indicators as a function of RMS of RVs. The vertical axis of (a), (b), (c) and (d) correspond to FWHM, $S_{\rm H{\alpha}}$, $V_{\rm span}$ and $W_{\rm span}$, respectively. Red and blue circles correspond to stars with $v\sin i$ of $\leq20\ {\rm km\,s^{-1}}$ and $\geq25\ {\rm km\,s^{-1}}$, respectively. The correlation coefficient and p-value in \% are shown in each panel at right top corner.}
\label{rvstat4}
\end{figure}

Figure \ref{rvstat4} shows the RMS values of activity indicators and RVs.
We can see that the RMS of FWHM, $V_{\rm span}$ and $W_{\rm span}$ show a significant correlation with that of RVs. 
This indicates that these three indicators are useful in evaluating the effect of stellar activities on RV variations.
\begin{eqnarray}
\label{eq_jitter2}
\sigma_{\rm FWHM} &=& 8.9\times10^{-3}(\sigma_{\rm RV})^{0.66}\\
\sigma_{V_{\rm span}} &=& 1.2\times10^{-2}(\sigma_{\rm RV})^{0.69}\\
\sigma_{W_{\rm span}} &=& 2.0\times10^{-1}(\sigma_{\rm RV})^{0.27}
\end{eqnarray}
where $\sigma_{\rm FWHM}$, $\sigma_{V_{\rm span}}$ and $\sigma_{W_{\rm span}}$ are the RMS of FWHM, $V_{\rm span}$ and $W_{\rm span}$ expressed in ${\rm km\,s^{-1}}$, respectively.

Among our sample, 17 stars were observed with photometry in Campaign field 4 of the K2 mission.
Cool spots on the stellar surface can not only cause spectral line-profile deformation but also change its brightness.
Therefore, we investigated the photometric data to see if there is a correlation between RV jitter and photometric modulation.
The photometry data are retrieved from MAST K2 Search Interface (https://archive.stsci.edu/k2/).
For each star, we fitted the photometric data with a 5th-order polynomial and normalized by using the fitted function to remove long-timescale variations.
As a result, we could not see any correlation between the two RMSs.
This can be due to the difference in observational epochs between our survey and the K2 mission.
While our survey was conducted between November 2017 and December 2017, Campaign field 4 of the K2 mission was conducted between February 2015 and April 2015.
Therefore, there is no guarantee that the stellar activity will remain the same over the two years.
The time span of observations may also be the cause.
The duration of our surveys was shorter ($\sim$15 day) than the K2 mission ($\sim$80 day).
\subsection{Planet occurrence rate}
Although the planet occurrence rate derived from our sample gave us a constraint on the HJs population around young stars, we attempted to enforce a more stringent constraint by combining our results with those obtained by other surveys targeting young clusters.
For this purpose, we referred to \authorcite{Paulson2006} (\yearcite{Paulson2006}, hereafter P06) and \authorcite{Bailey2018} (\yearcite{Bailey2018}, hereafter B18).
P06 observed four young open clusters and moving groups by the RV method; $\beta$ Pic association ($\sim$12 Myr), IC 2391 (30--50 Myr), Castor moving group (200$\pm$100 Myr) and Ursa Major moving group (300 Myr). 
We do not refer to $\beta$ Pic association due to its poor constraints on the detection limit.
B18 derived the detection limits for 117 stars in NGC 2516 (141 Myr) and NGC 2422 (73 Myr) based on RV observations.
Since P06 showed the detection limits for planets with orbital periods less than 6 days and B18 showed those with orbital periods of 3, 10, and 20 days, we only considered the detection limit for a 3-days period for consistency.
The fact that the detection limit of P06 for the 3-days period may be smaller than that derived for the 6-days one makes no significant impact on our discussion because at the 6-days period, their detection limits are low enough to enable the investigation of the HJs with masses larger than $1\ M_{\rm JUP}$.
Among our sample, the corotation radius, namely, the inner edge of a disk at which migrating planets will halt, lies inside the location of a 3-days orbital period, and thus the occurrence rate of short-period planets should give us insight into the formation process of HJs.

\begin{figure}
\begin{center}
\includegraphics[width=10cm]{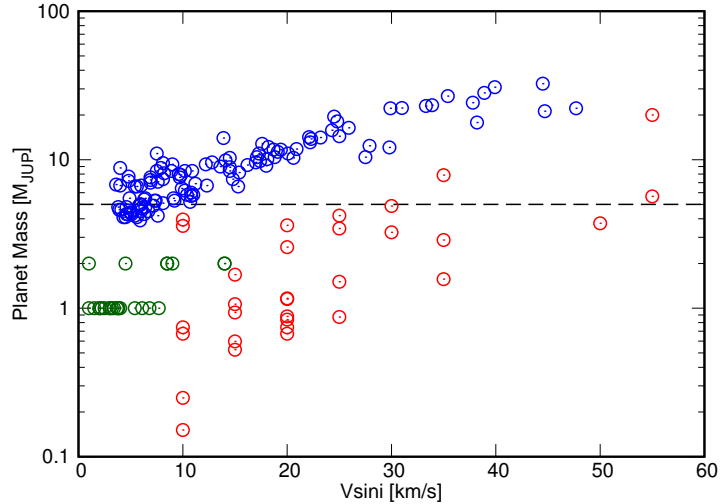}
\end{center}
\caption{The planet mass detection limit at 3-days period as a function of $v\sin i$. Red, blue, and green circles represent the detection limit derived by this work, B18 and P06, respectively. The dashed line is drawn at $5\ M_{\rm JUP}$.}
\label{freq1}
\end{figure}

Figure \ref{freq1} shows the detection limits derived by this work, B18 and P06, respectively.
As seen in the figure, B18 derived a higher detection limit compared to ours.
Since the ages of open clusters observed by our survey and B18 are almost identical, the stellar RV jitter should be almost the same.
The discrepancy might be explained by the difference in the V magnitude of sample stars.
While the V magnitude of our sample ranges from 9.5 to 10.5 mag, B18 observed stars with 12--16 magnitude.
Furthermore, while we improved the detection limits by correcting stellar activities, B18 did not use any such techniques.
As for $v\sin i$, most of the stars observed by P06 have rotational velocities less than $20\ {\rm km\,s^{-1}}$.
This simply comes from target selection criteria.
P06 set a rotational velocity criterion of $v\sin i<20\ {\rm km\,s^{-1}}$.
We note that our surveys achieved detection limits down to 5 $M_{\rm JUP}$ for stars with various rotational velocities (from 10 to 55 ${\rm km\,s^{-1}}$).

Hereafter in order to evaluate the difference of the planet-detection sensitivities among the three surveys, we divide the mass range into three: 1--5, 5--13 and 13--80 $M_{\rm JUP}$.
We also limit the rotational velocity to $v\sin i<30\ {\rm km\,s^{-1}}$ for mass ranges of 1--5 and 5--13 $M_{\rm JUP}$.
Table \ref{tab_occ2} shows the planet occurrence rates that we derived with equations (4)--(7) following \citet{Borgniet2017}.
It should be noted that no planets were detected from the three surveys, and thus we focus on 1 $\sigma$ upper limit of the occurrence rate for the following discussion.

For the mass range of 1--5 $M_{\rm JUP}$, the occurrence rates derived with the result of this work or P06 alone are estimated to be less than 15.5\% and 12.6\% within 1 $\sigma$ error, respectively.
By combining the two surveys, the occurrence rate was improved to less than 7.3\% within 1 $\sigma$ error even though the small number of samples (50 stars) gave poor constraints on the occurrence rate.
However, it is difficult to compare this rate with that of field stars ($1.2\pm0.38\%$; \cite{Wright2012}).
In addition, the planet occurrence rate in open clusters is confronted with two conflicting claims.
\citet{Quinn2014} derived the occurrence rate of $1.97^{+1.92}_{-1.07}\%$ by combining their surveys (1 HJ among 26 stars in Hyades and 2 HJs among 60 stars in Praesepe) with the result of \citet{Paulson2004} (no HJ among 74 FGK stars in Hyades).
Since Hyades and Praesepe are relatively metal-rich ([Fe/H]$\sim+0.15$), the authors adjusted the value to match the solar-metallicity and derived $0.99^{+0.96}_{-0.54}\%$, which is consistent with \citet{Wright2012}.
On the other hand, \citet{Brucalassi2017} found the overabundance of HJs in open cluster M 67 ($\sim5.7^{+5.5}_{-3.0}\%$) because of encounters with other stars in the cluster and binary companions.
\citet{Shara2016} also showed that HJs can be produced via the encounters if the binary fraction in the cluster is large, and so is the case for M 67 (the binary fraction of 30\%; \cite{Latham2007}).
In contrast, HJs cannot be produced efficiently via the encounters if the binary fraction is small ($\sim10\%$).
Since the binary fraction in the Pleiades is small (4.9--6.1\%, \cite{Bouvier1997}; 5.9\%, \cite{Richichi2012}; 10.0\%, \cite{Konishi2016}), we expect that the cluster environment does not enhance the HJ formation in the Pleiades.
According to the N-body simulations conducted by \citet{Fujii2019}, close-in planets within 1 au can be rarely ejected in open cluster environment.
This indicates that the occurrence rate of HJs is retained after formation of them.
Therefore, if DDM is the main mechanism by which HJs form, the occurrence rate of HJs at the age of $\sim$100 Myr should be identical to that of the field stars.

For the mass range of 5--13 $M_{\rm JUP}$, B18 accounted for a large fraction of the samples and gave the occurrence rate of 4.9\% as an upper limit.
As shown in table \ref{tab_occ2}, the completeness of B18 in this mass range is lower compared to those of this work and P06.
By combining the samples of this work and P06 with those of B18 in order to improve the completeness, the occurrence rate is estimated to be less than 2.9\% within 1 $\sigma$ error.
From transit surveys, the paucity of massive HJs (MHJ) around solar-type stars has been claimed (e.g., \cite{Deleuil2008}).
\citet{Bouchy2011} suggested that MHJ around G-type dwarf stars would be engulfed by loss of angular momentum via magnetic braking.
\citet{Guillot2014} confirmed \citet{Bouchy2011}'s suggestion by using more sophisticated numerical simulations and predicted the lifetime of the planets with orbital periods of 3 days around various stars (0.8--1.4 $M_{\odot}$). 
For example, planets with masses of 5--13 $M_{\rm JUP}$ orbiting solar-mass stars can survive for several Gyr.
Therefore, MHJs should still be surviving at the age of $\sim$100 Myr. 
The lack of detection of MHJs in our survey, B18, and P06 might indicate that such planets are originally less likely to be formed around solar-type stars, or that the tidal interaction between the stars and their planets are stronger than considered in the theoretical models, which results in a shorter lifetime of close-in massive planets.
A caveat is that the observed paucity of the detection of the MHJs is greater than that predicted by \citet{Guillot2014} and their lifetime might be much shorter.
\citet{Damiani2016} suggest that the ignorance of tidal dissipation in the stellar convective envelope could cause the difference between the theoretical prediction and the practical observations.

Finally, we determined the occurrence rate for close-in brown dwarfs (BD) in a mass range of 13--80 $M_{\rm JUP}$ to be less than 1.9\%.
As seen in table \ref{tab_occ2}, the search completenesses in this mass range are close to 100\% for each survey.
Therefore, we can confidently rule out the existence of close-in BDs with orbital periods of $\sim$3 days.
Similarly to the case of the MHJs, the close-in BDs are preferentially detected around F-type stars by the transit survey.
\citet{Damiani2016} performed numerical simulations to investigate the orbital evolution of close-in BDs around F-, G-, K- and M-type stars.
Their results showed that BDs with orbital periods less than 3 days around G5 stars can live only for several hundred Myr.
Therefore the lack of detection of close-in BDs in our survey, B18, and P06 is in agreement with theoretical predictions.

\begin{table}
\tbl{Planet occurrence rates derived by this work, P06, and B18.\footnotemark[$*$]}{%
 \begin{tabular}{ccccccccc}
 \hline\hline
Mass & Survey	&Number of stars& $v\sin i$ &Completeness & Missed planets& Occurrence rate& 1$\sigma$ error& 2$\sigma$ error\\
$(M_{\rm JUP})$& 		& 		&$({\rm km\,s^{-1}})$ &(\%)& &(\%)&(\%) & (\%)\\\hline
1--5 	&1+2	&	50	&$<$30	&	87.4 	&	0.14 	&	2.3 	&	7.2	&	10.3\\
1--5	&	1	&	24	&$<$30	&	81.1	&	0.23	&	5.2	&	15.5	&	21.7	\\
1--5	&	2	&	26	&$<$30	&	93.2	&	0.07	&	4.1	&	12.6	&	17.6	\\\hline
5--13 & 1+2+3 	&	156 	& $<$30	&	73.3	&	0.36	&	0.8	&	2.9	&	4.1\\
5--13	&	1	&	24	&$<$30	&	94.7	&	0.06	&	4.4	&	13.3	&	18.6	\\
5--13	&	2	&	26	&$<$30	&	100	&	0.00	&	3.8	&	11.7	&	16.4	\\
5--13	&	3	&	106	&$<$30	&	62.0	&	0.61	&	1.5	&	4.9	&	7.0	\\\hline
13--80 & 1+2+3	&	173 	& 	All 	& 	98.5 & 	0.02 	& 	0.6	&	1.9	&	2.7\\ 
13--80 &	1	&	30	&	All	&	99.3	&	0.01	&	3.3	&	10.3	&	14.5	\\
13--80 &	2	&	26	&	All	&	100	&	0.00	&	3.8	&	11.7	&	16.4	\\
13--80 &	3	&	117	&	All	&	97.9	&	0.02	&	0.9	&	2.8	&	4.0	\\\hline
 \end{tabular}}
\begin{tabnote}
 \hangindent6pt\noindent
\hbox to6pt{\footnotemark[$*$]\hss}\unskip%
Column 1: mass range. Column 2: surveys used to derive the planet occurrence rate. 1, 2, and 3 correspond to this work, P06 and B18, respectively. Column 3: the number of sample stars. Column 4: the limit on $v\sin i$. Column 5--7: the search completeness, number of missed planets, and planet occurrence rate, respectively
\end{tabnote}
\label{tab_occ2}
\end{table}

\subsection{HJs formation process}
The purpose of our survey is to put a constraint on the HJs formation process.
Given the different time-scale between DDM and HEM, the frequency of HJs around stars at the age of less than $\sim$ 100 Myr should provide us with important clues.
If DDM is a major process, HJs formation is completed by that age.
This implies that the occurrence rate of HJs for young stars should be comparable to that for field stars ($\sim1\%$).
On the other hand, if HEM is a major process, it takes several hundred Myr to form HJs.
Hence, HJs should be less frequent.
 
Currently, no HJs were detected in our survey as well as B18 and P06.
This null detection sets an upper limit on the HJs frequency.
Our sample set an upper limit for the occurrence rate of HJs to be 11.4\% (Table. 3).
\citet{Heller2019} predicted an upper limit of the initial HJ formation rate via DDM to be 41\%.
Disk lifetimes suggest that a large fraction of HJs formed via DDM may spiral into their host stars by the first 10 Myr and some of the rest would do later via tidal interactions between planets and host stars.
The lower frequency of HJs compared to the upper limit of the initial HJ formation rate suggests that the infall of HJs into their stars have almost finished by the age of the Pleiades open cluster ($\sim$100 Myr).
On the other hand, the HJs frequency that we derived is higher than that of field stars.
This would imply that the tidally-induced infall of HJs takes longer than 100 Myr.
A more precise frequency of HJs helps us to understand the tidal interactions between planets and host stars.

Our upper limit on the HJs frequency is not constrained well enough to distinguish two major formation scenarios (i.e., DDM and HEM) of HJs because of insufficient sample size.
Base on binomial statistics, we need approximately 330 samples to confidentially determine the occurrence rate to be less than 1\% in the case of no planet detection.
Therefore, a larger sample is required to distinguish the HJs formation process.
\section{Summary}
We conducted a radial-velocity search for short-period planets in the Pleiades open cluster to investigate their formation and evolution process.
The observations were done between November and December 2017 at Okayama Astrophysical observatory with High Dispersion Echelle Spectrograph.
We obtained 3--9 spectra for each of our 30 sample stars with the typical SNR of 30--70.

Firstly, we computed the RVs for our sample and found that the RV measurement error was worse than expected.
Although we have not yet fully clarified the cause of degradation, the wavelength-dependent stellar activities might provide a partial explanation.
We also computed four stellar activity indicators.
Several stars showed significant RV variations compared to RV measurement errors.
By comparing RVs with the activity indicators, we found strong correlations between them and concluded that these RV variations were caused by stellar activities.
From our observations, no planet candidates were detected.

Secondly, we performed a statistical analysis on our sample.
The detection limits helped us to check the existence of planets.
By correcting the effect of stellar activities on RVs, the detection limits were improved by a factor of 2--3.
Then, combining the detection limits of the whole sample, we derived the search completeness for our survey.
We divided our sample into two sub-samples based on stellar rotational velocities.
As a result, we could rule out the existence of short-period planets with masses of 3 $M_{\rm JUP}$ and 10 $M_{\rm JUP}$, among the sub-sample of slowly and rapidly rotating stars, respectively.
We then derived the planet occurrence rate in the Pleiades open cluster, which was the first attempt to give a constraint on the occurrence rate of planets in the Pleiades.
Our results showed that the occurrence rate of planets with masses of 1--13 $M_{\rm JUP}$ and orbital periods of 1--10 days is less than 11.4\% within 1 $\sigma$ error.

By combining our survey with two other surveys targeting the open clusters in the age groups of 30--300 Myr, we attempted to enforce a more stringent constraint on the planet occurrence rate with an orbital period of 3 days.
To take the difference between the detection sensitivities of each survey into account, we divided the planet masses into three ranges; 1--5, 5--13, and 13--80 $M_{\rm JUP}$. 
As for the mass range of 1--5 $M_{\rm JUP}$, we determined the occurrence rate to be less than 7.2\% in 1$\sigma$ error.
As for the mass range of 5--13 and 13--80 $M_{\rm JUP}$, we found the occurrence rate to be less than 2.9\% and 1.9\% within 1 $\sigma$ error, respectively.
Therefore, the lack of detection of massive short-period planets might indicate that such planets are originally less likely to be formed around solar-type stars.

\bigskip
\begin{ack}
This research is based on data collected at the Okayama Astrophysical Observatory (OAO), which is operated by National Astronomical Observatory of Japan. 
We are grateful to all the staff members of OAO for their support during the observations.
This work is supported by the Astrobiology Center Program of National Institutes of National Sciences (NINS) (Grant Number JY300109).
\end{ack}

\appendix
\section*{Results of various analysis for our sample} \label{appendix1}
\begin{figure}[htb]
\begin{center}
\includegraphics[width=1.0\textwidth]{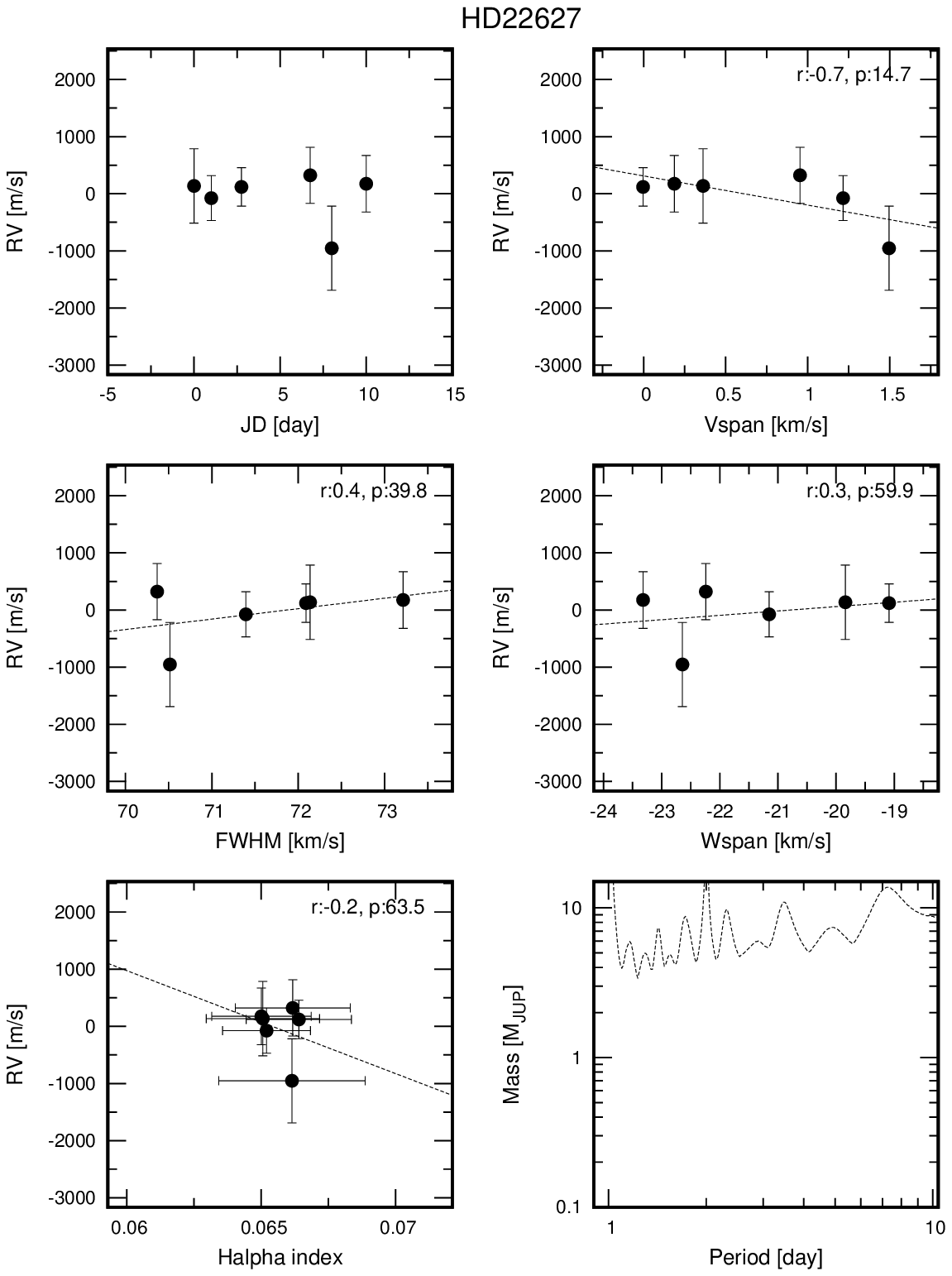}
\end{center}
\vspace{-0.7cm}
\caption{The results of various analyses for HD 22627. Left column: RV vs JD, FWHM, and S$_{\rm H_{\alpha}}$ (from top to bottom). 
The first observational date is subtracted from JD.
Right column: RV vs V$_{\rm SPAN}$, W$_{\rm SPAN}$, and the detection limit (from top to bottom).
In the panel of the detection limit, the dashed and solid lines are the results before and after correcting stellar activities, respectively.
As for the panels of RV vs FWHM, S$_{\rm H_{\alpha}}$, V$_{\rm SPAN}$ and W$_{\rm SPAN}$, the fitted linear functions are represented by dotted lines, and the correlation coefficients and p-value (in \%) are shown at top-right corner.}
\label{appen1}
\end{figure}

\newpage
\begin{figure}[htb]
\begin{center}
\includegraphics[width=1.0\textwidth]{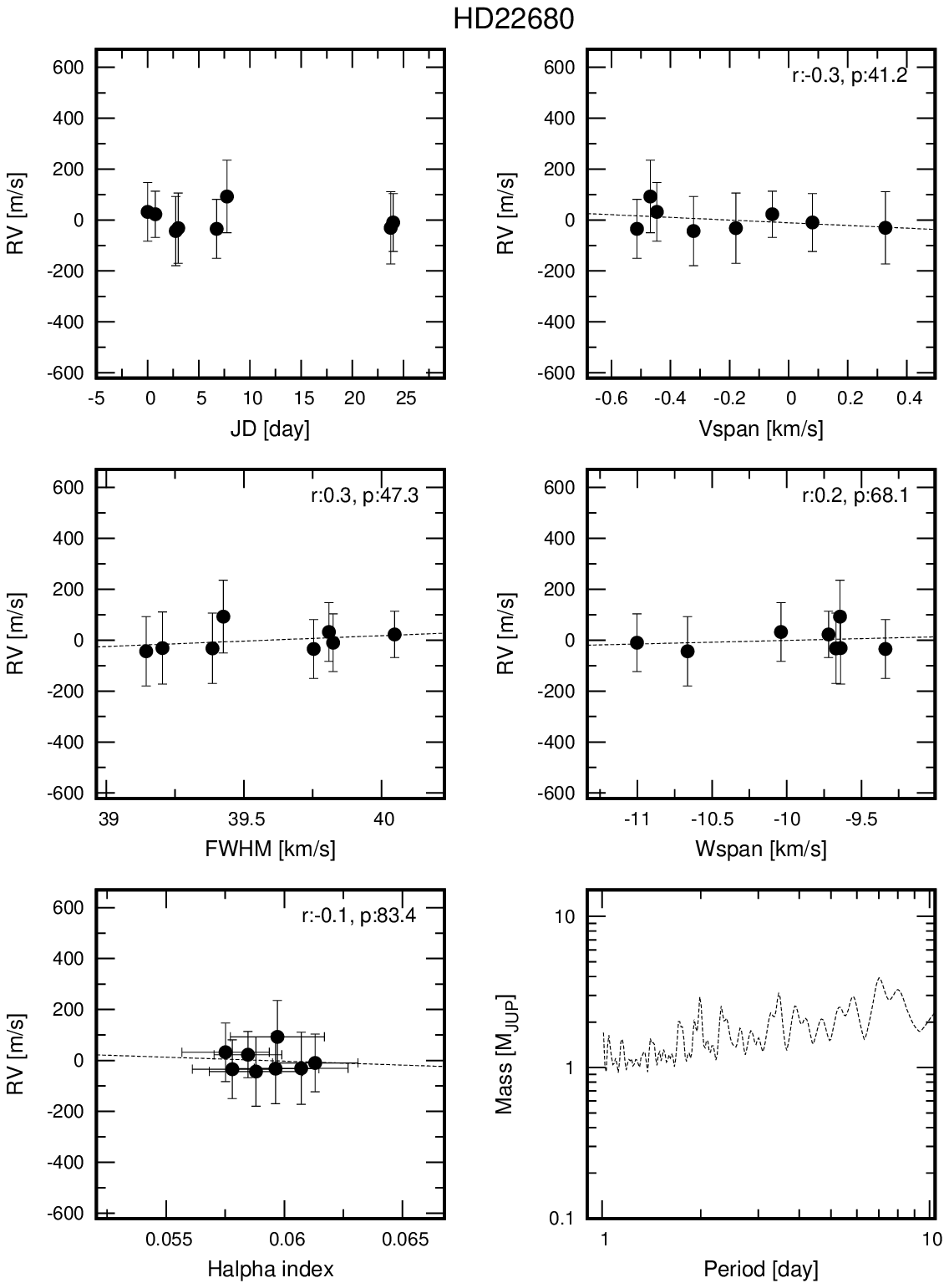}
\end{center}
\vspace{-0.7cm}
\caption{Same as figure \ref{appen1} for HD 22680.}
\end{figure}

\newpage
\begin{figure}[htb]
\begin{center}
\includegraphics[width=1.0\textwidth]{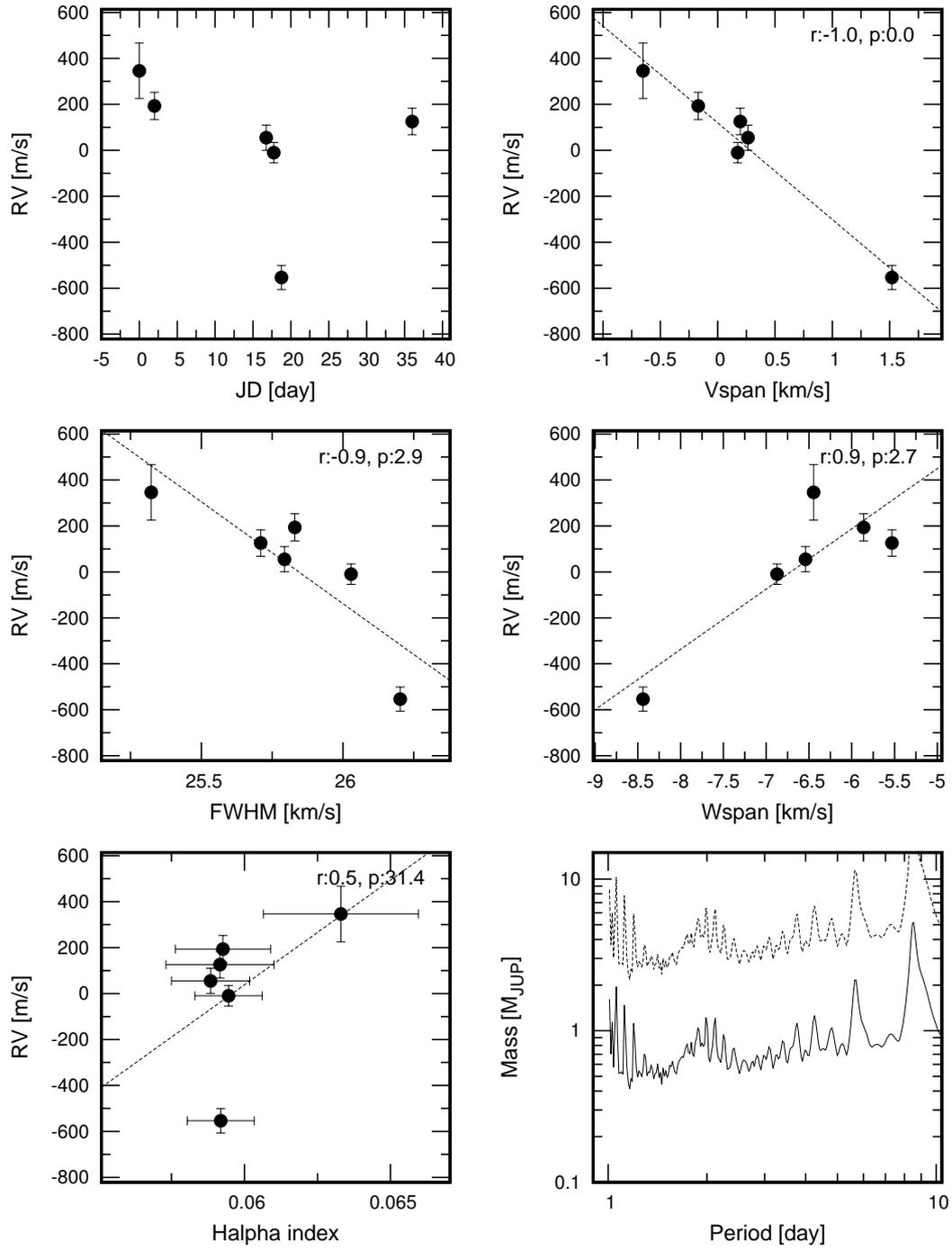}
\end{center}
\vspace{-0.7cm}
\caption{Same as figure \ref{appen1} for HD 23269.}
\end{figure}

\newpage
\begin{figure}[htb]
\begin{center}
\includegraphics[width=1.0\textwidth]{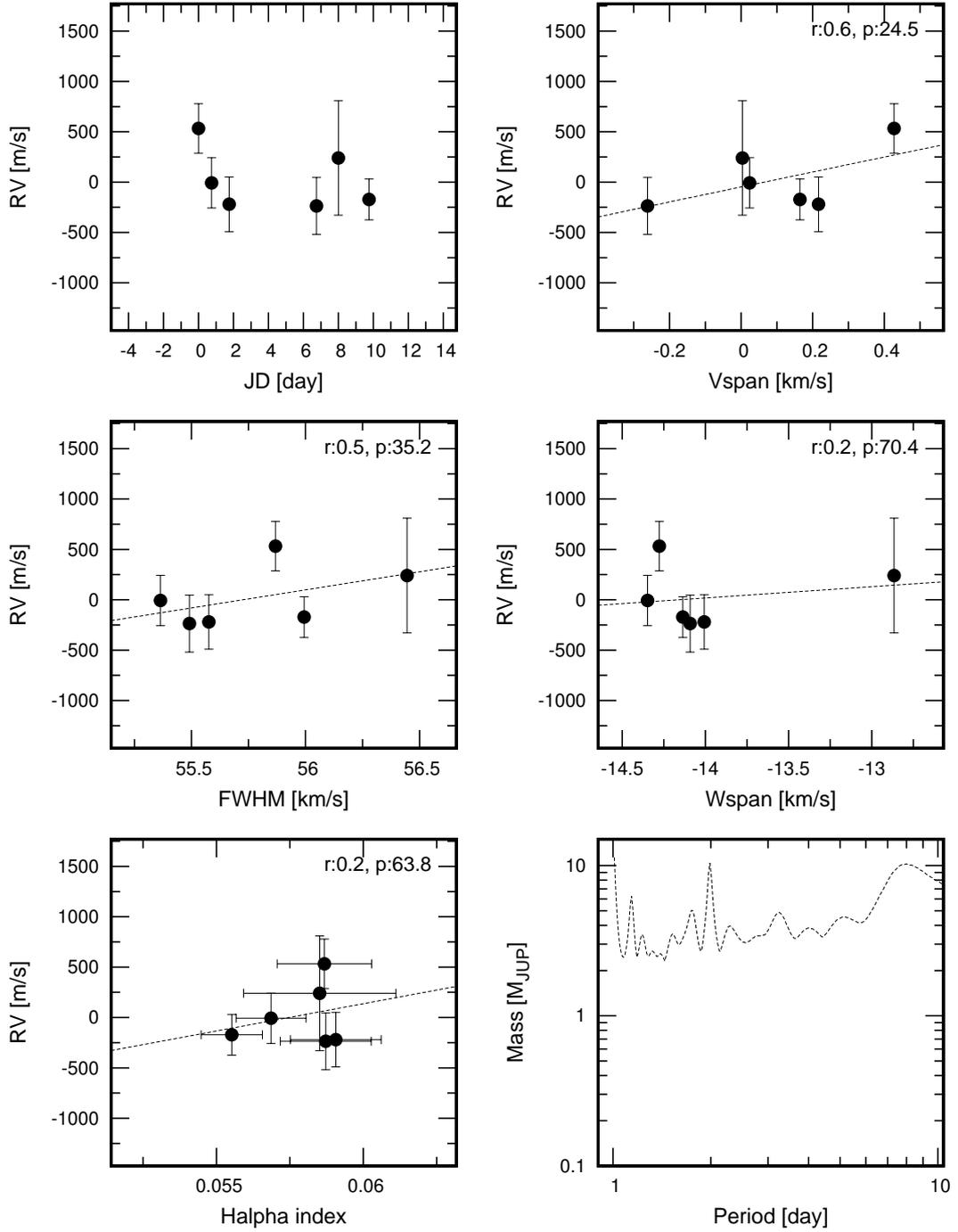}
\end{center}
\vspace{-0.7cm}
\caption{Same as figure \ref{appen1} for HD 23514.}
\end{figure}

\newpage
\begin{figure}[htb]
\begin{center}
\includegraphics[width=1.0\textwidth]{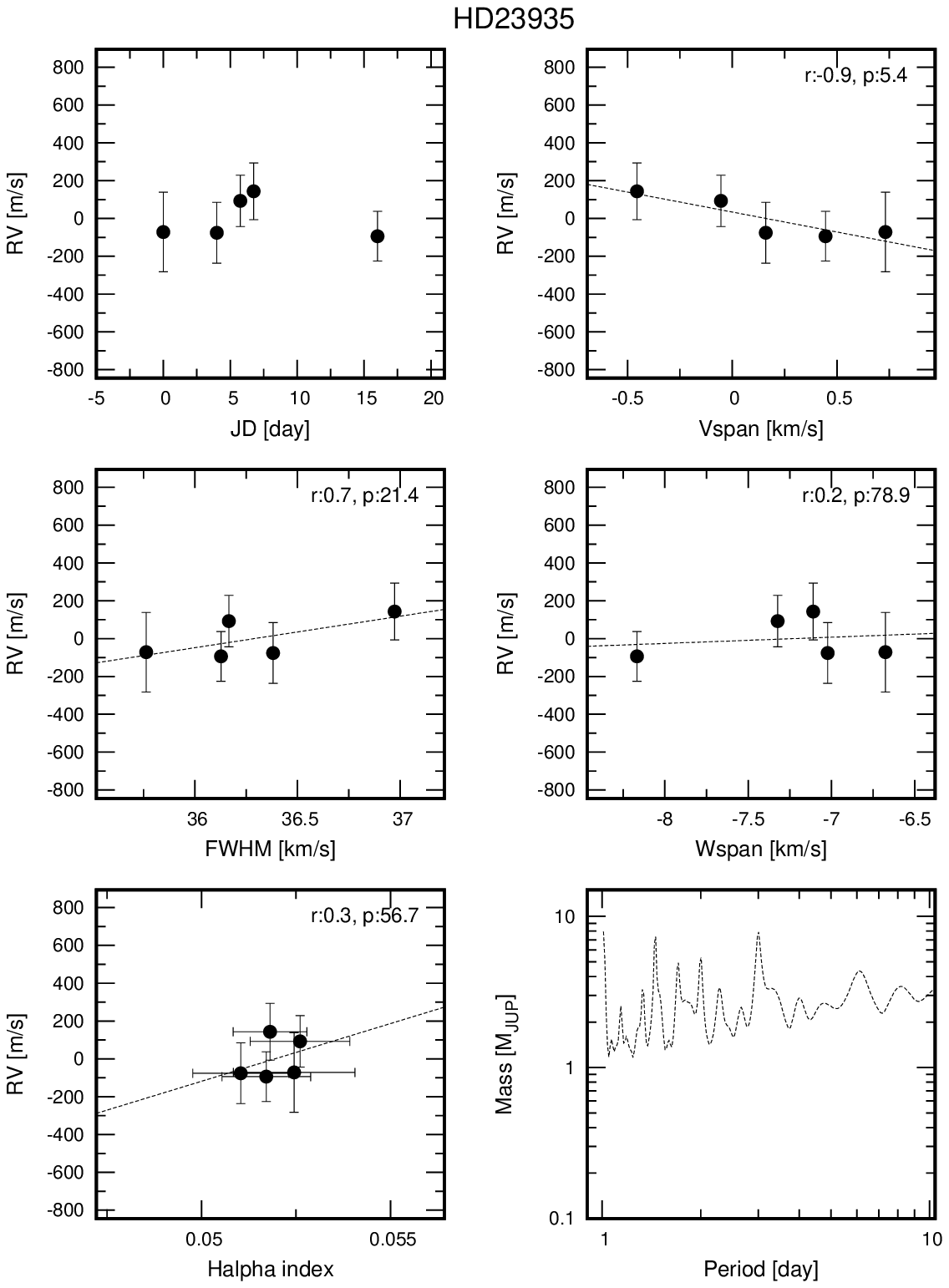}
\end{center}
\vspace{-0.7cm}
\caption{Same as figure \ref{appen1} for HD 23935.}
\end{figure}

\newpage
\begin{figure}[htb]
\begin{center}
\includegraphics[width=1.0\textwidth]{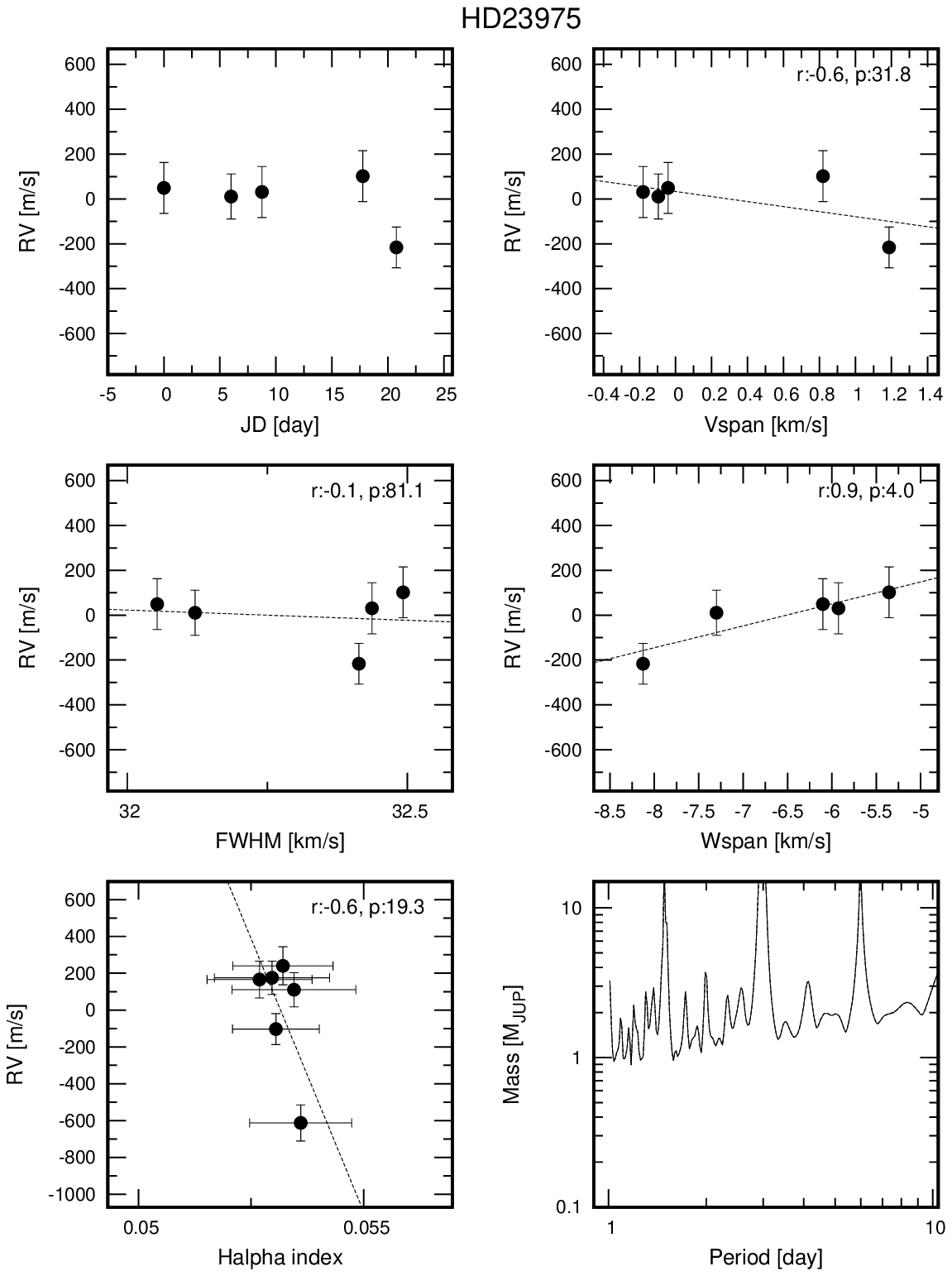}
\end{center}
\vspace{-0.7cm}
\caption{Same as figure \ref{appen1} for HD 23975.}
\end{figure}

\newpage
\begin{figure}[htb]
\begin{center}
\includegraphics[width=1.0\textwidth]{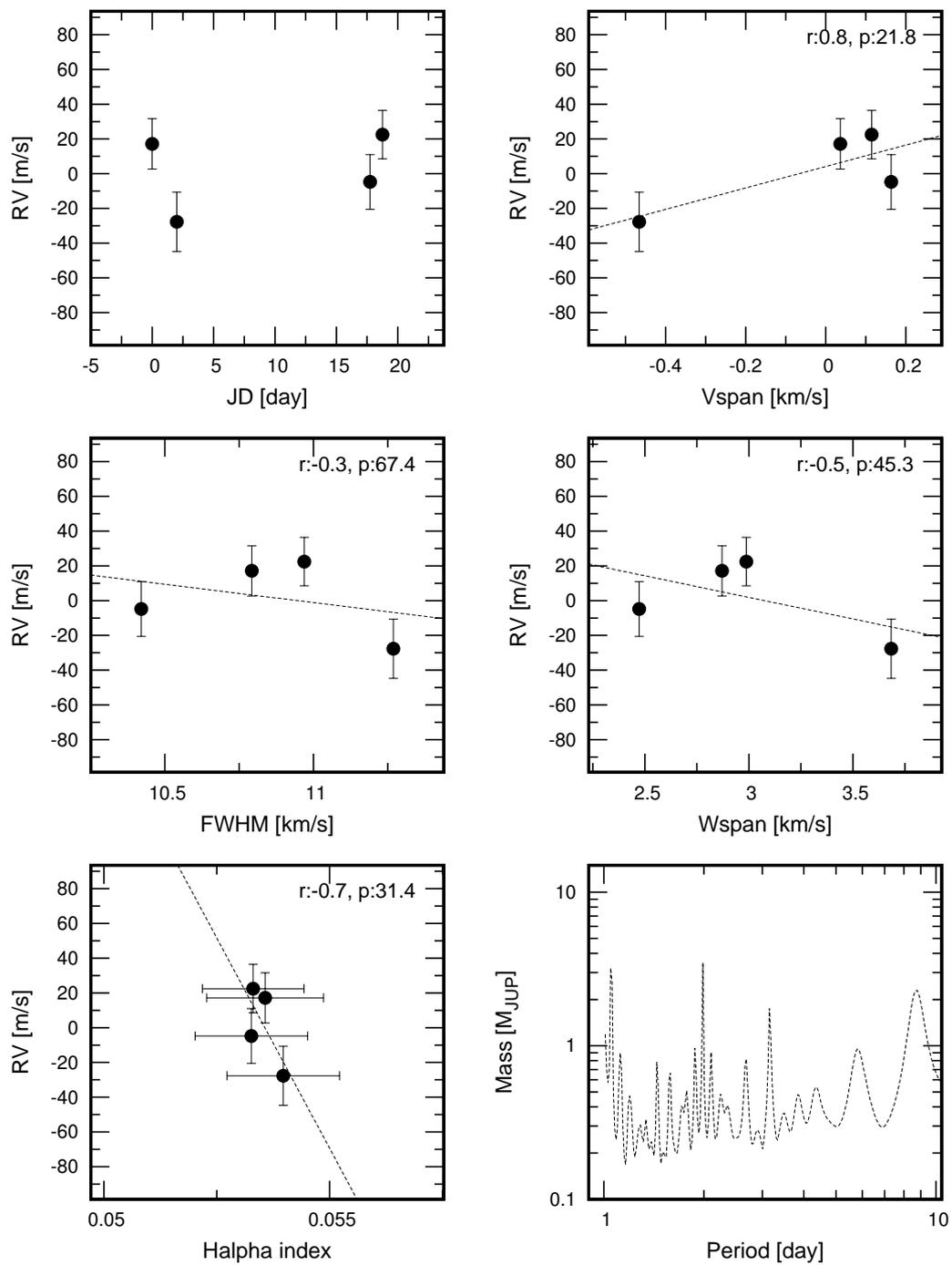}
\end{center}
\vspace{-0.7cm}
\caption{Same as figure \ref{appen1} for HD 24194.}
\end{figure}

\newpage
\begin{figure}[htb]
\begin{center}
\includegraphics[width=1.0\textwidth]{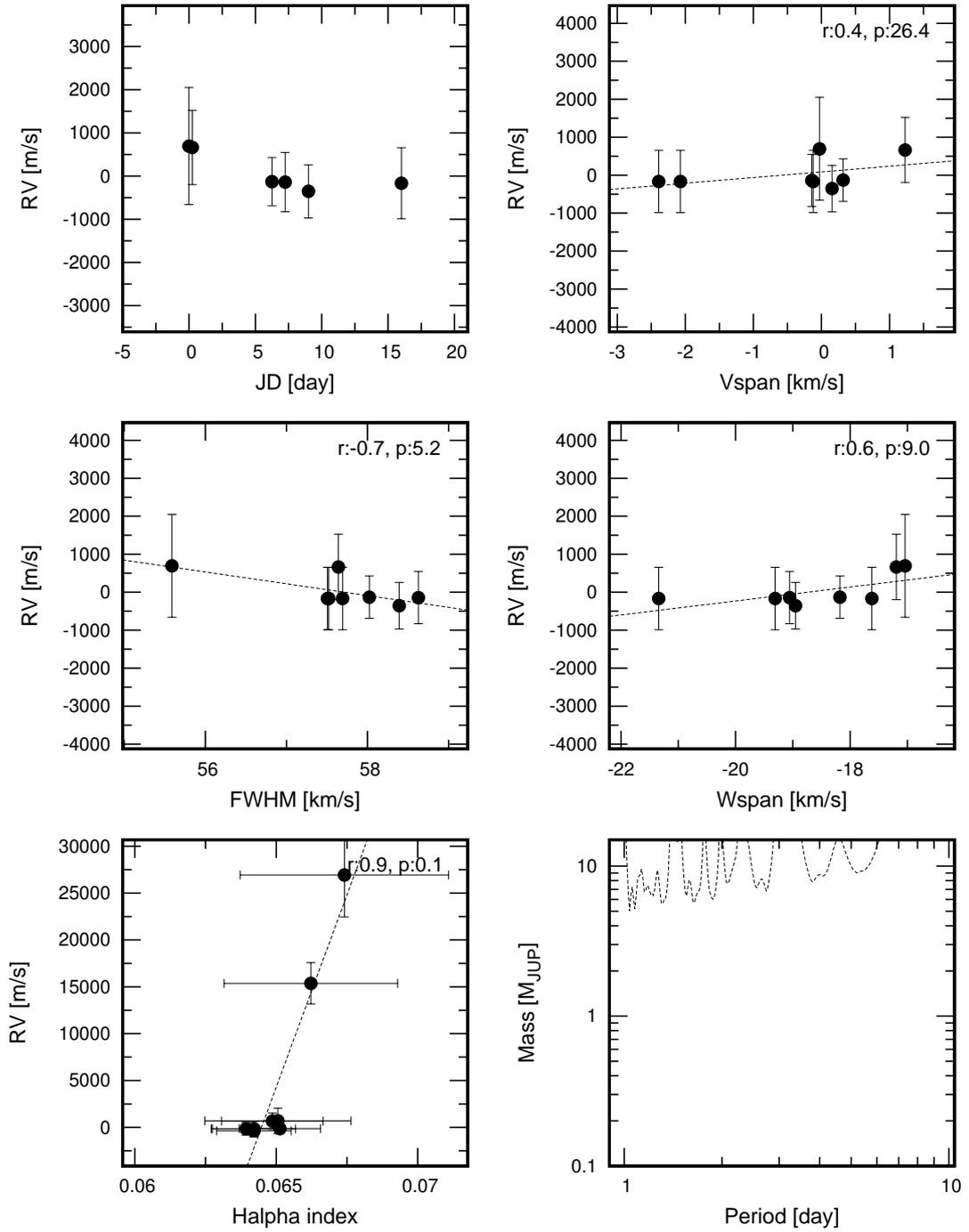}
\end{center}
\vspace{-0.7cm}
\caption{Same as figure \ref{appen1} for HD 24302.}
\end{figure}

\newpage
\begin{figure}[htb]
\begin{center}
\includegraphics[width=1.0\textwidth]{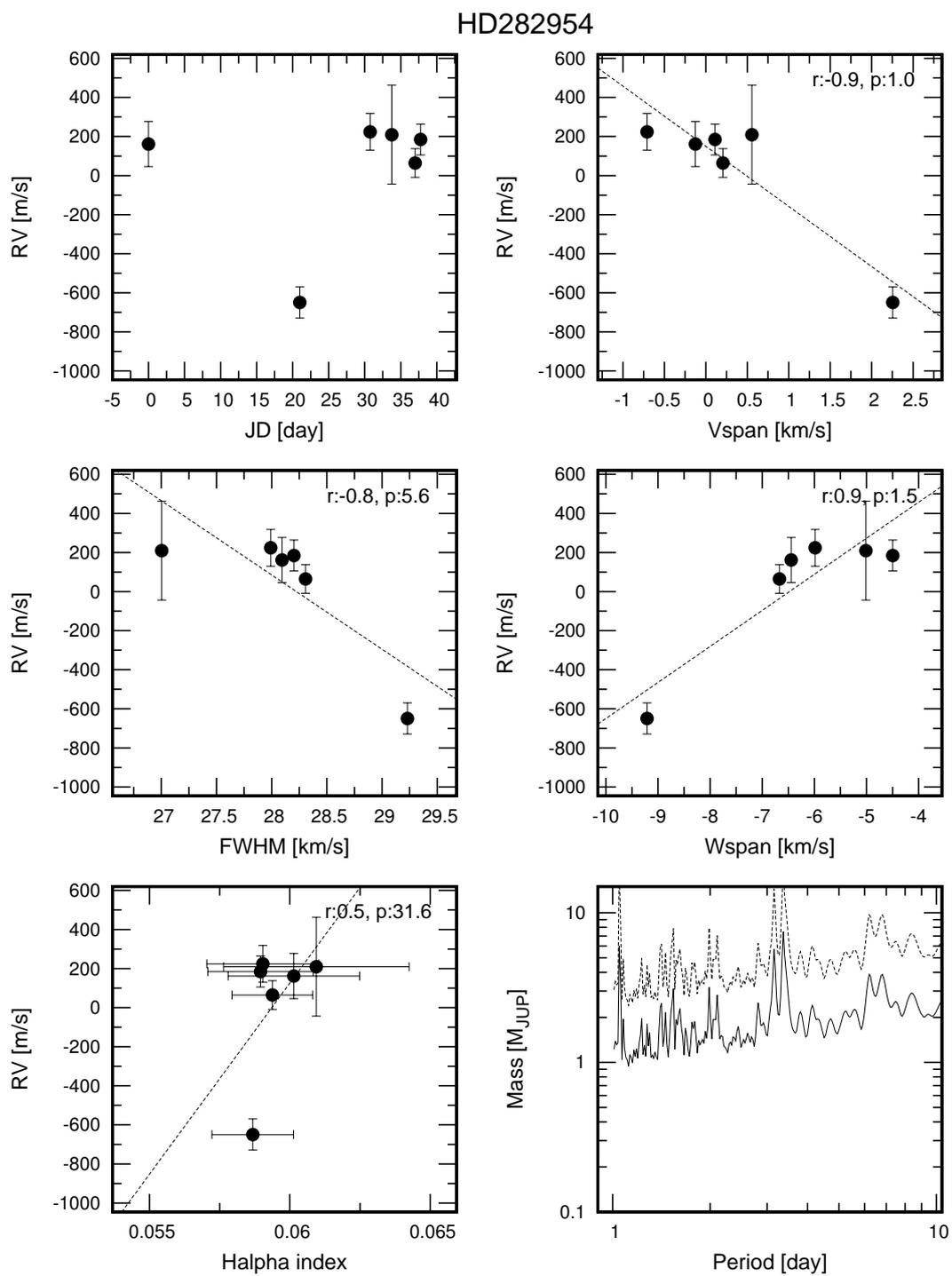}
\end{center}
\vspace{-0.7cm}
\caption{Same as figure \ref{appen1} for HD 282954.}
\end{figure}

\newpage
\begin{figure}[htb]
\begin{center}
\includegraphics[width=1.0\textwidth]{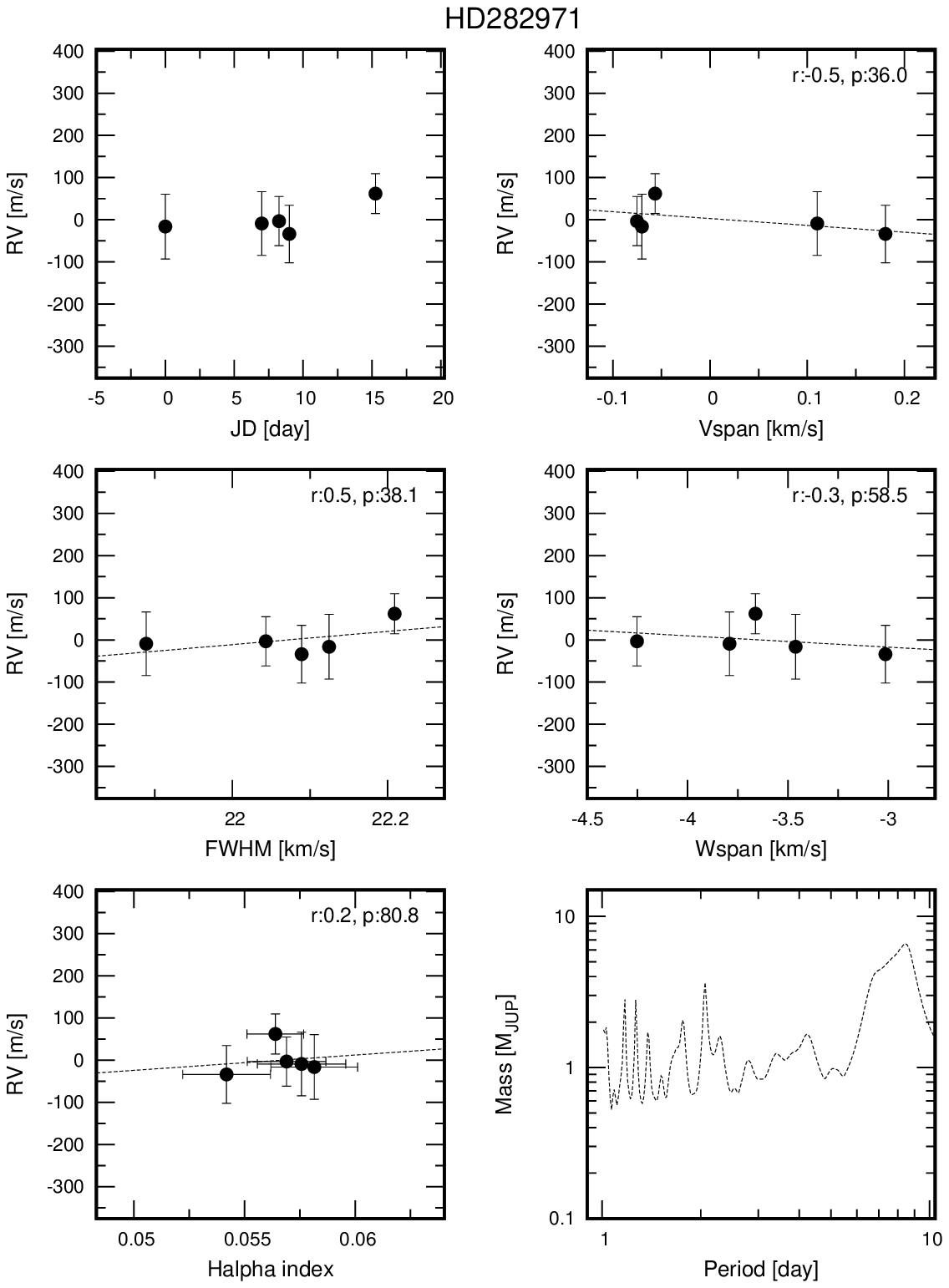}
\end{center}
\vspace{-0.7cm}
\caption{Same as figure \ref{appen1} for HD 282971.}
\end{figure}

\newpage
\begin{figure}[htb]
\begin{center}
\includegraphics[width=1.0\textwidth]{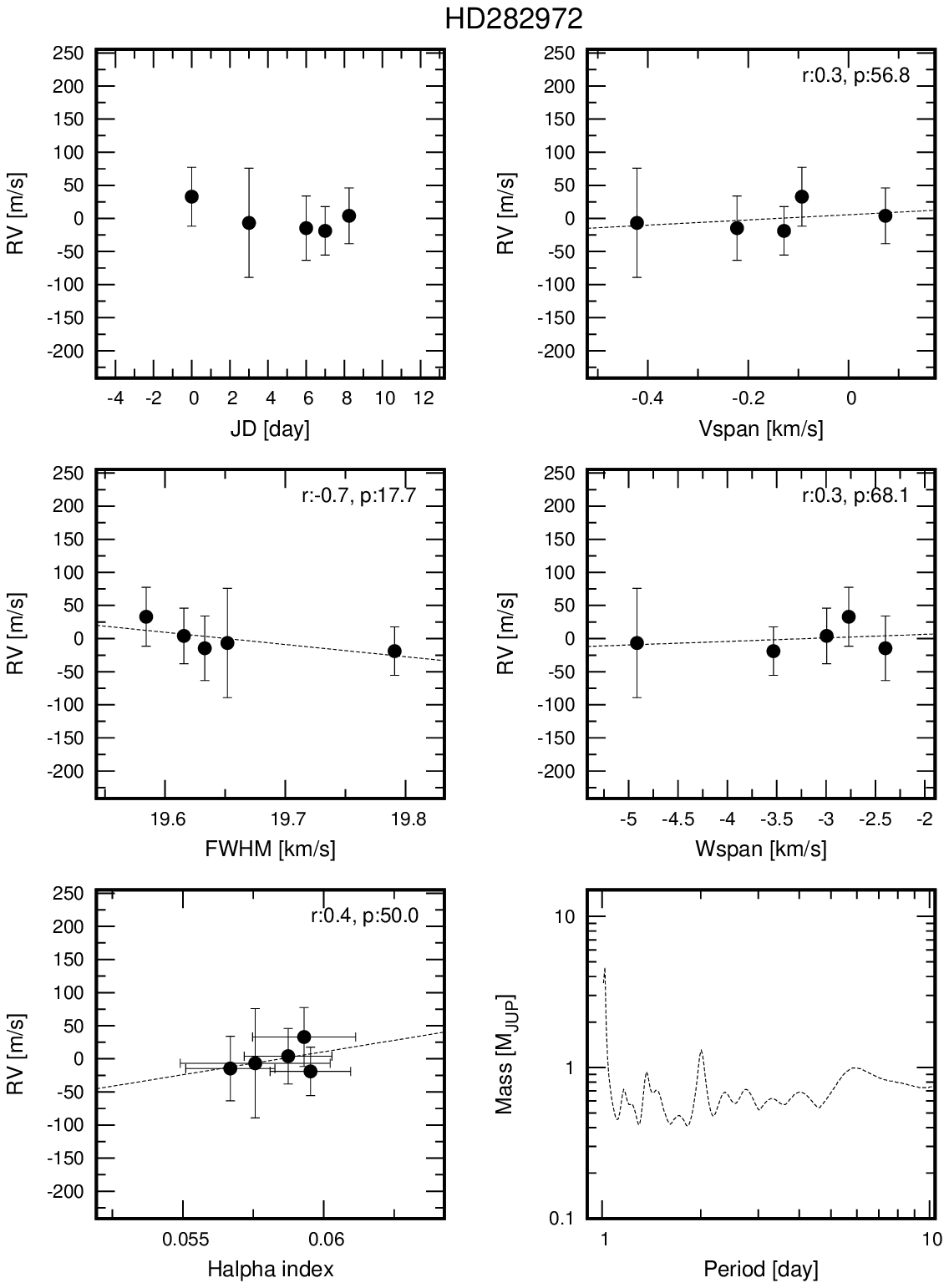}
\end{center}
\vspace{-0.7cm}
\caption{Same as figure \ref{appen1} for HD 282972.}
\end{figure}

\newpage
\begin{figure}[htb]
\begin{center}
\includegraphics[width=1.0\textwidth]{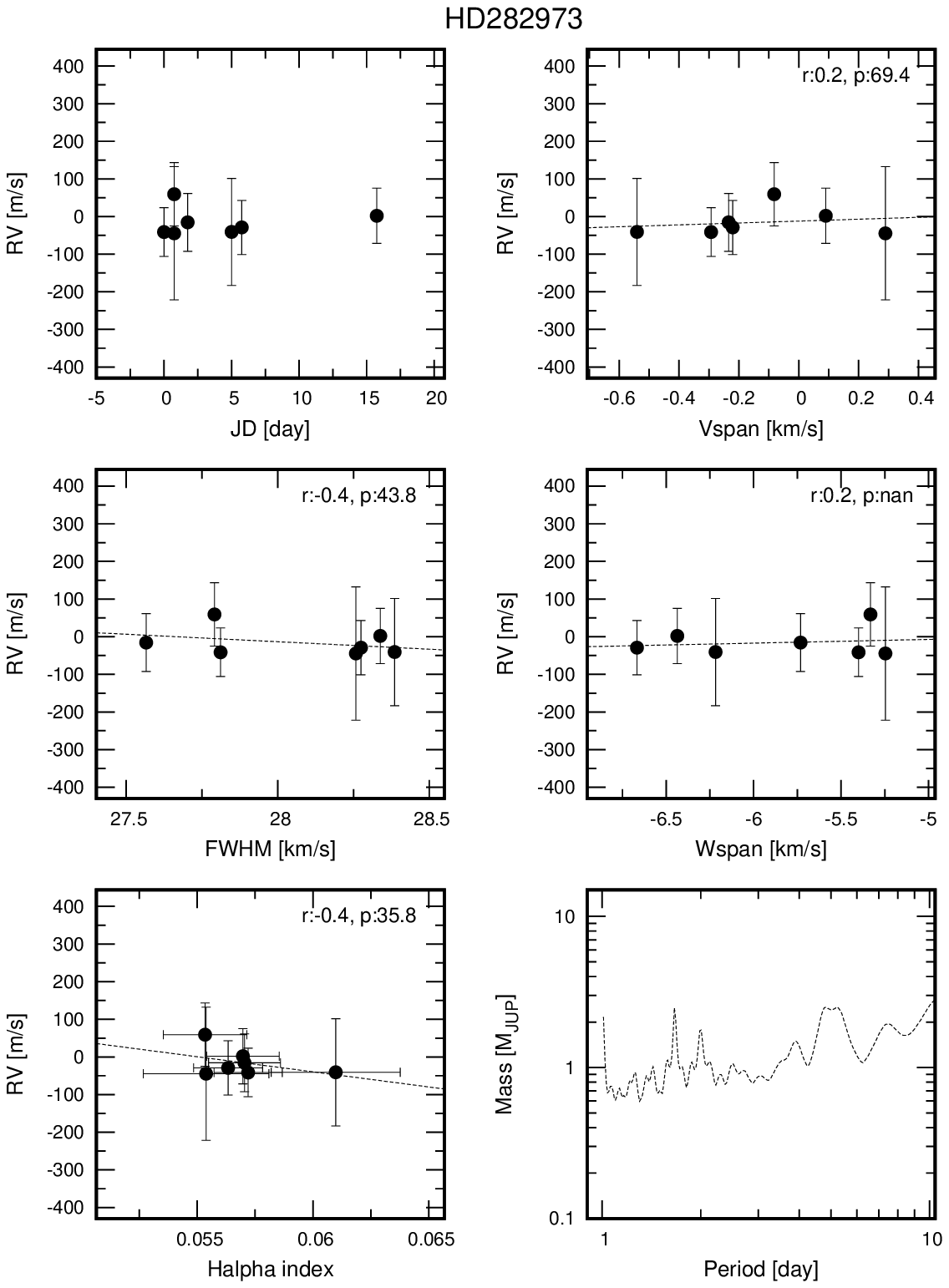}
\end{center}
\vspace{-0.7cm}
\caption{Same as figure \ref{appen1} for HD 282973.}
\end{figure}

\newpage
\begin{figure}[htb]
\begin{center}
\includegraphics[width=1.0\textwidth]{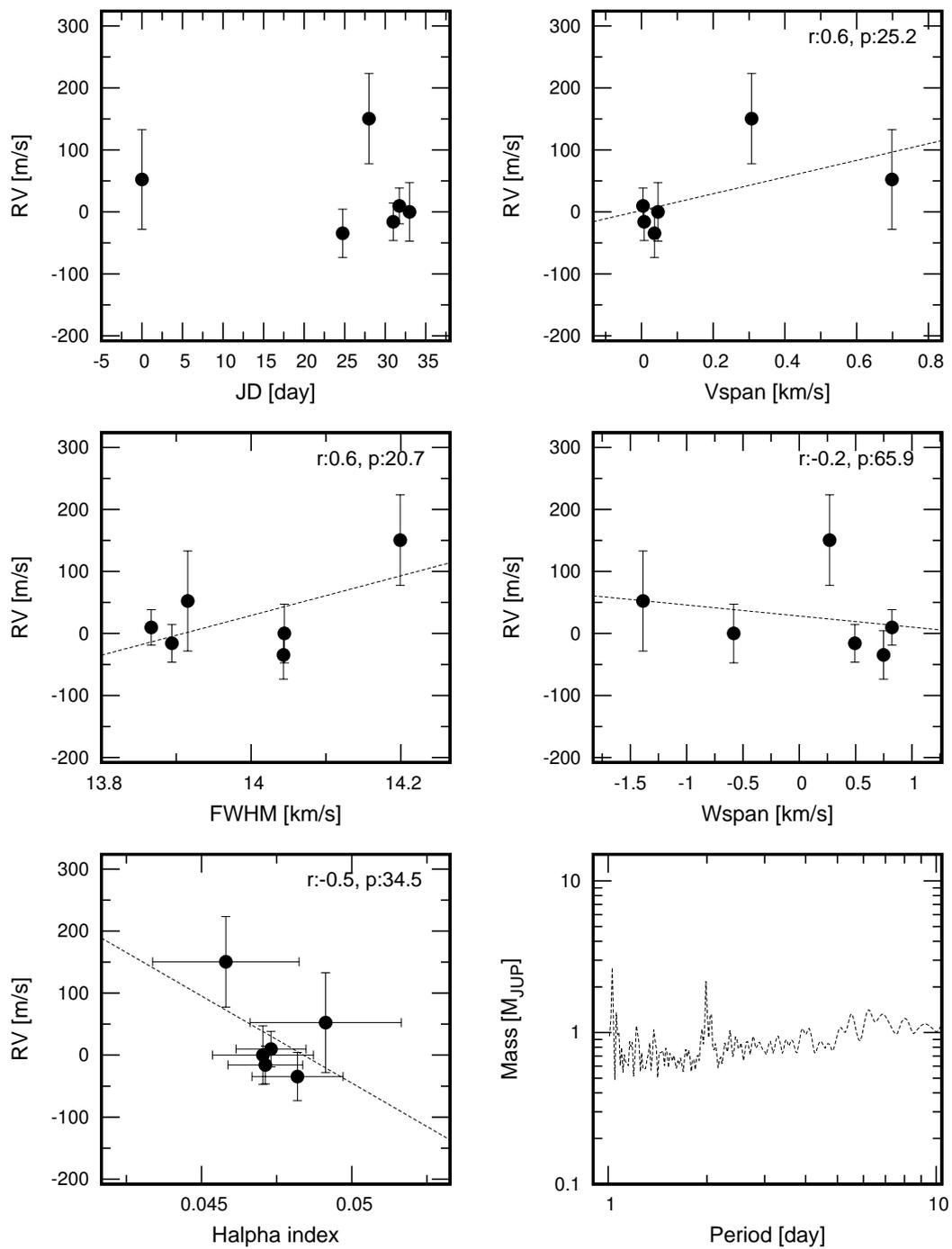}
\end{center}
\vspace{-0.7cm}
\caption{Same as figure \ref{appen1} for HD 282998.}
\end{figure}

\newpage
\begin{figure}[htb]
\begin{center}
\includegraphics[width=1.0\textwidth]{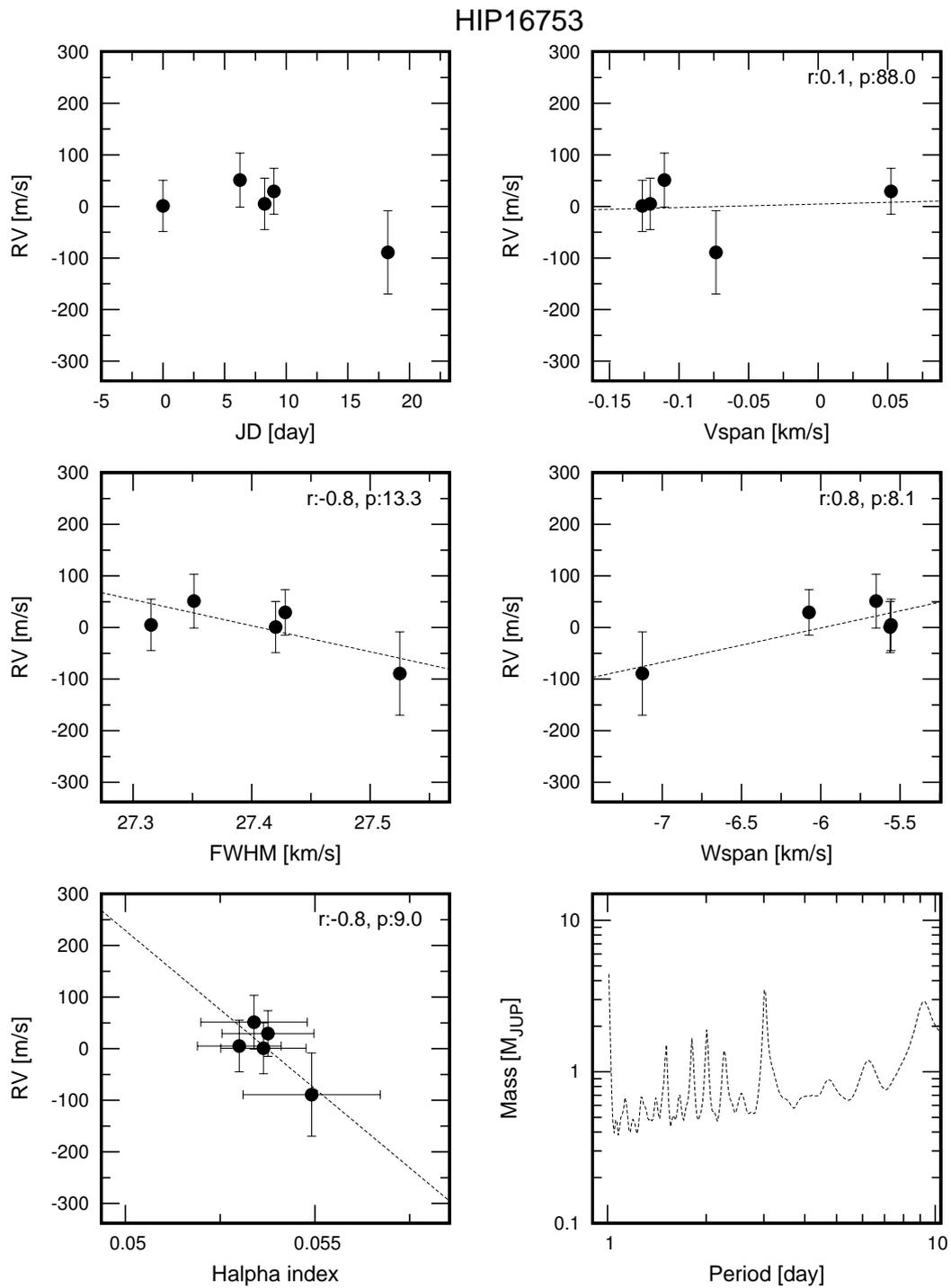}
\end{center}
\vspace{-0.7cm}
\caption{Same as figure \ref{appen1} for HIP 16753.}
\end{figure}

\newpage
\begin{figure}[htb]
\begin{center}
\includegraphics[width=1.0\textwidth]{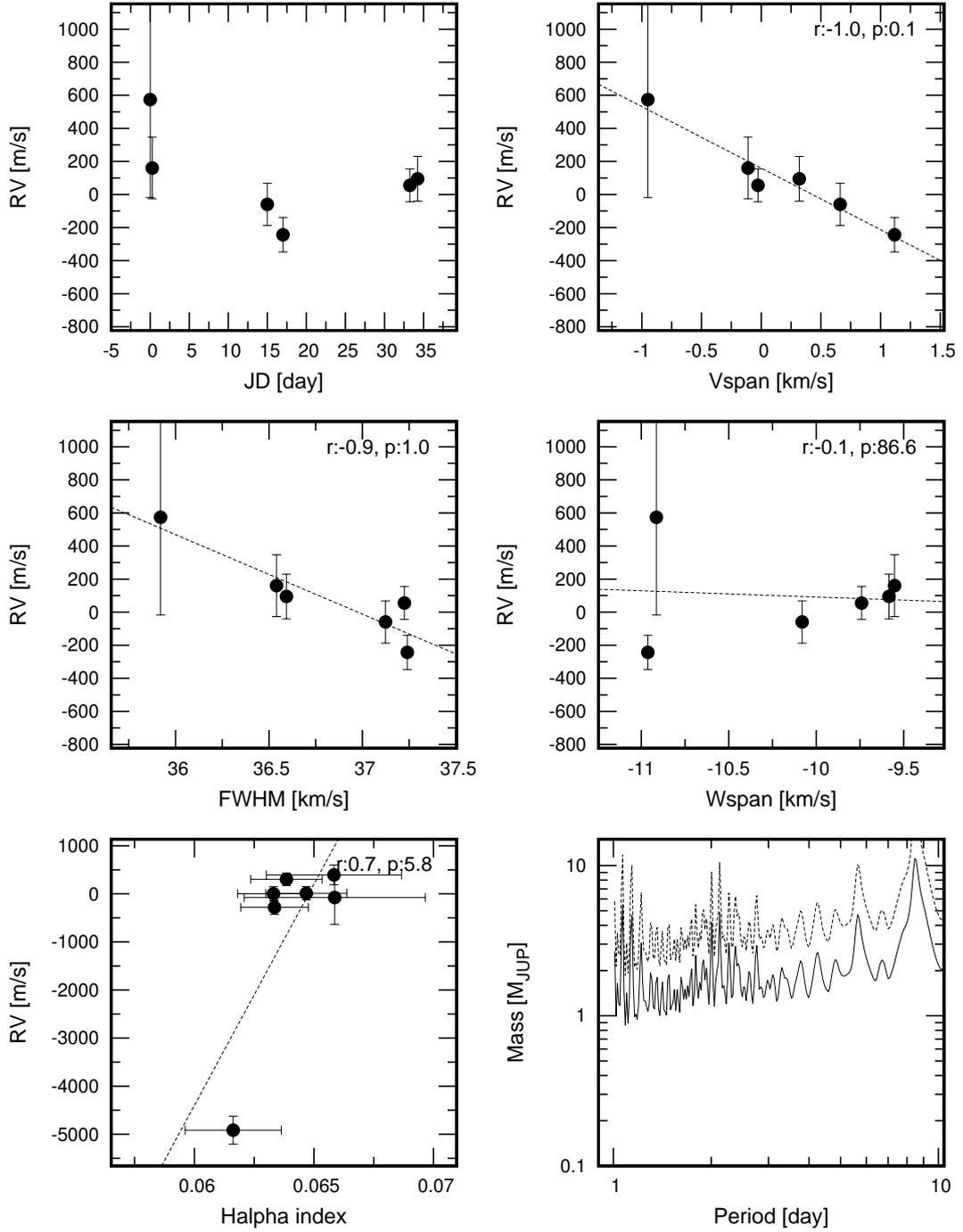}
\end{center}
\vspace{-0.7cm}
\caption{Same as figure \ref{appen1} for HIP 16979.}
\end{figure}

\newpage
\begin{figure}[htb]
\begin{center}
\includegraphics[width=1.0\textwidth]{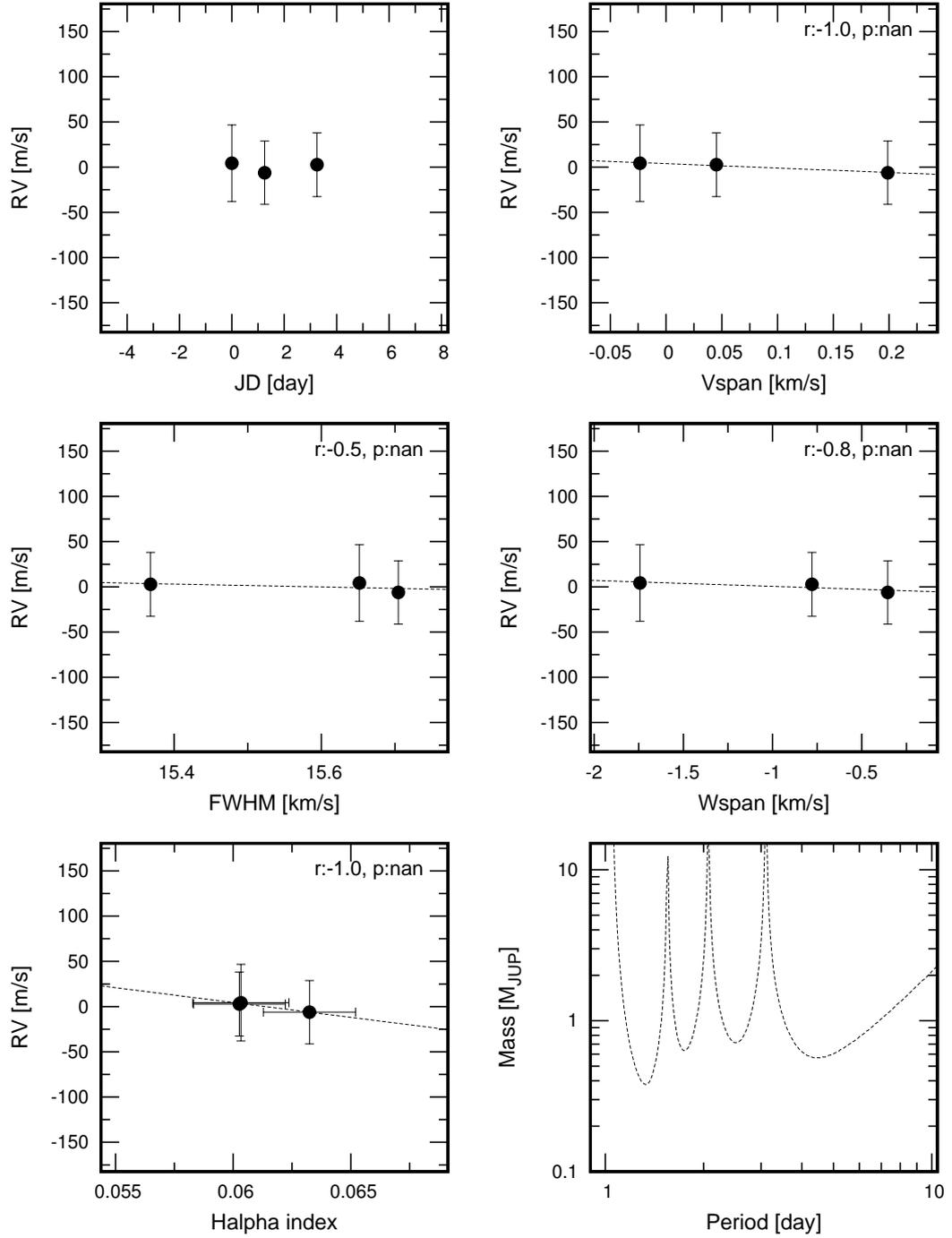}
\end{center}
\vspace{-0.7cm}
\caption{Same as figure \ref{appen1} for HIP 17020.}
\end{figure}

\newpage
\begin{figure}[htb]
\begin{center}
\includegraphics[width=1.0\textwidth]{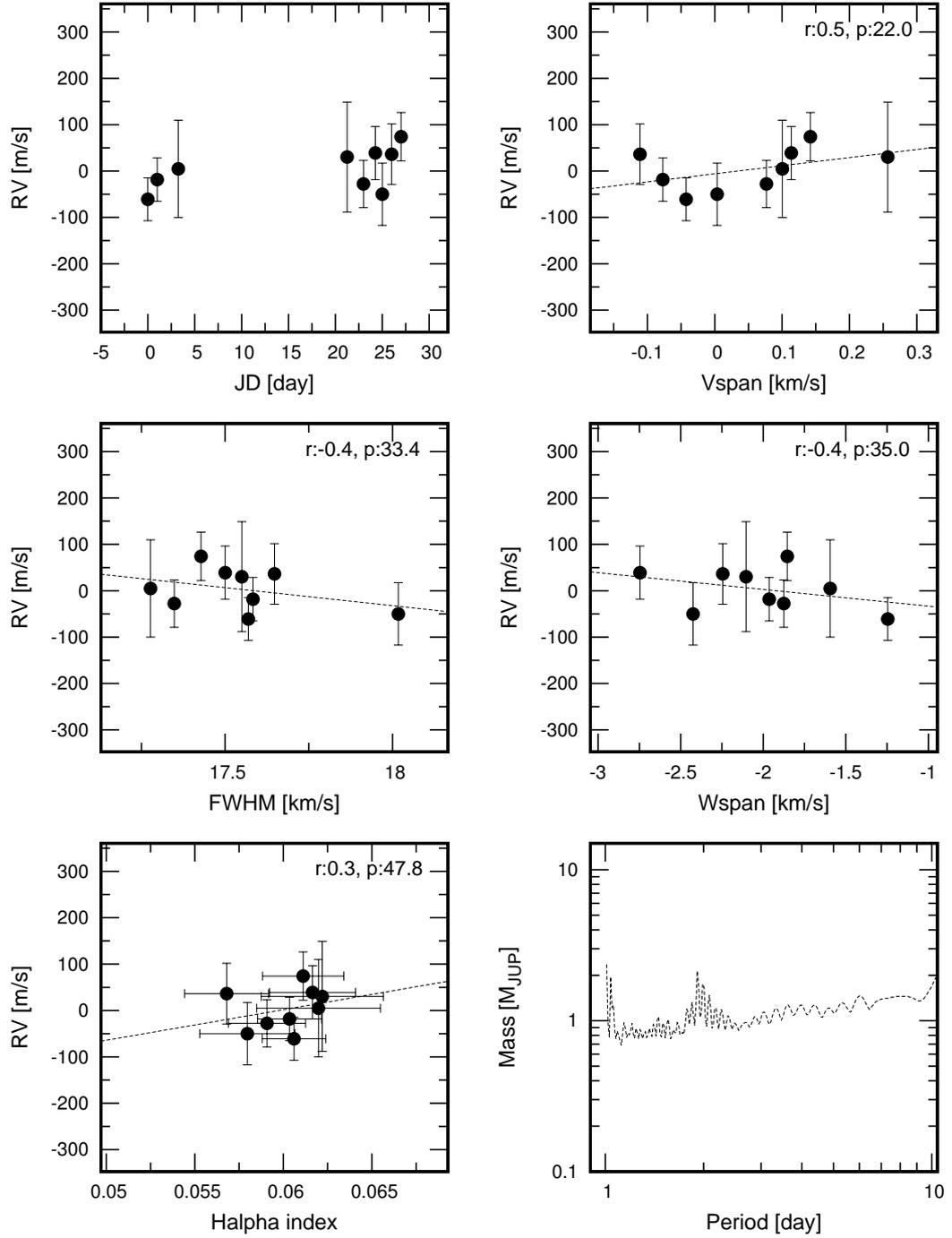}
\end{center}
\vspace{-0.7cm}
\caption{Same as figure \ref{appen1} for HIP 17044.}
\end{figure}

\newpage
\begin{figure}[htb]
\begin{center}
\includegraphics[width=1.0\textwidth]{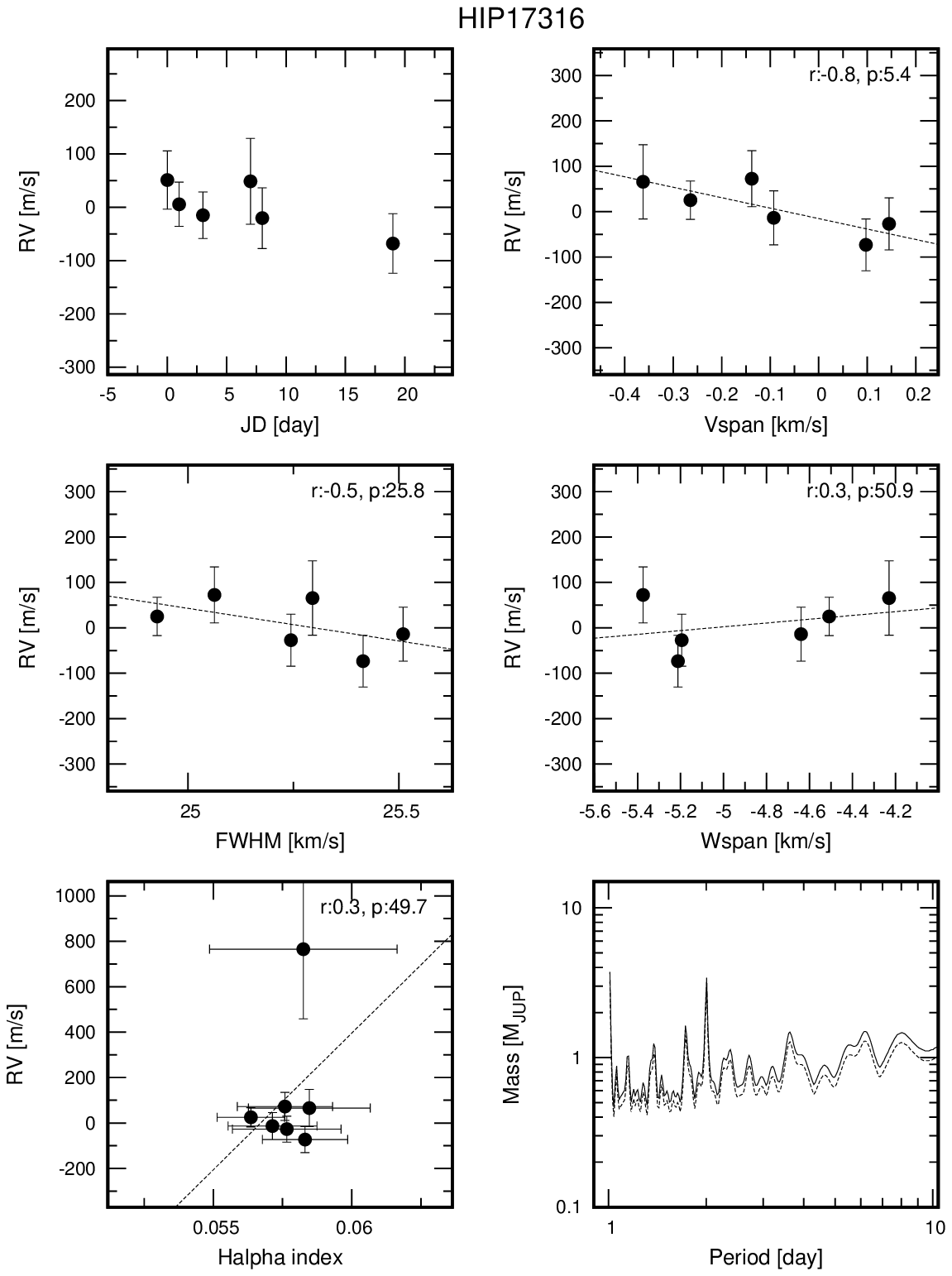}
\end{center}
\vspace{-0.7cm}
\caption{Same as figure \ref{appen1} for HIP 17316.}
\end{figure}

\newpage
\begin{figure}[htb]
\begin{center}
\includegraphics[width=1.0\textwidth]{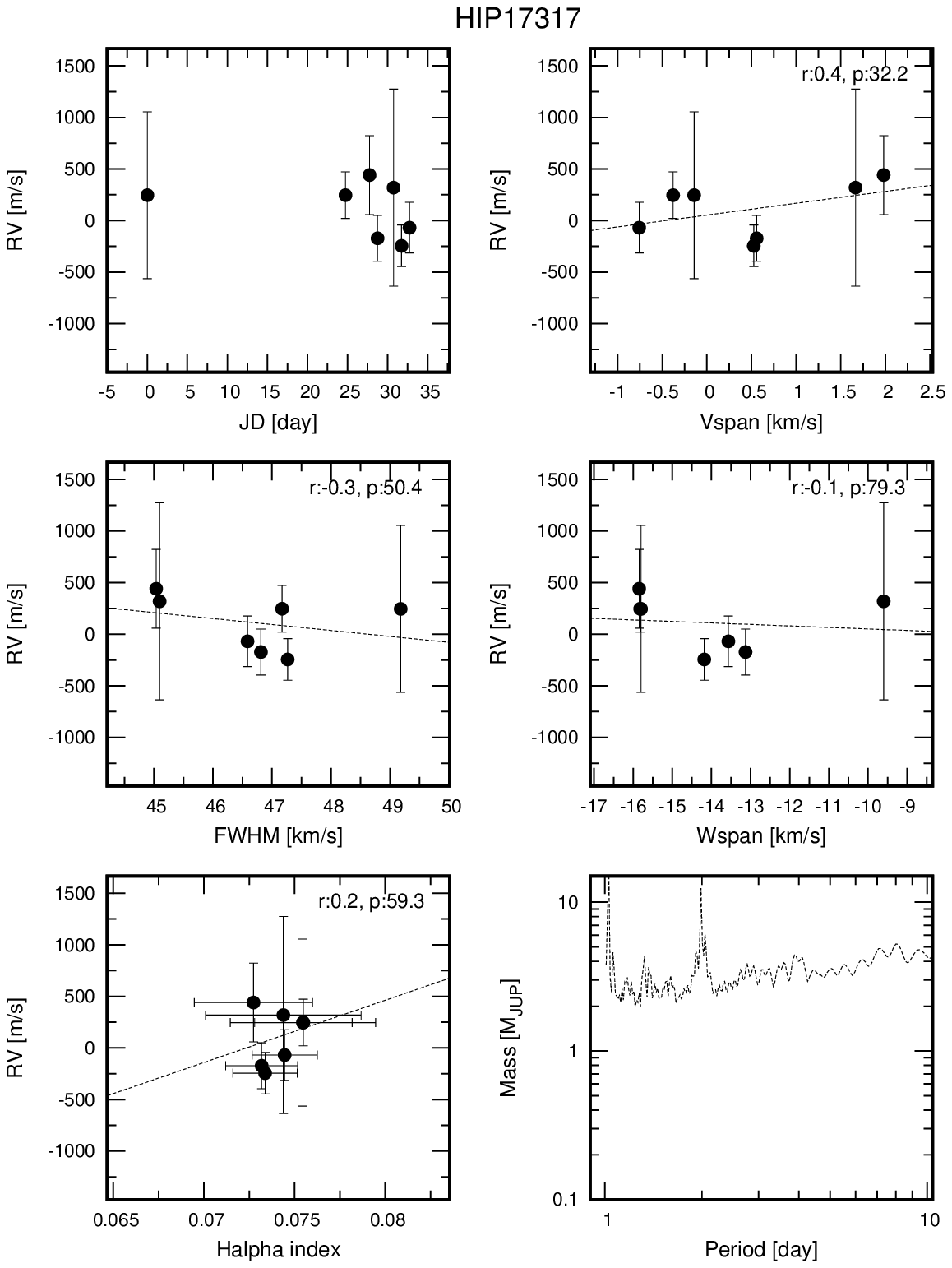}
\end{center}
\vspace{-0.7cm}
\caption{Same as figure \ref{appen1} for HIP 17317.}
\end{figure}

\newpage
\begin{figure}[htb]
\begin{center}
\includegraphics[width=1.0\textwidth]{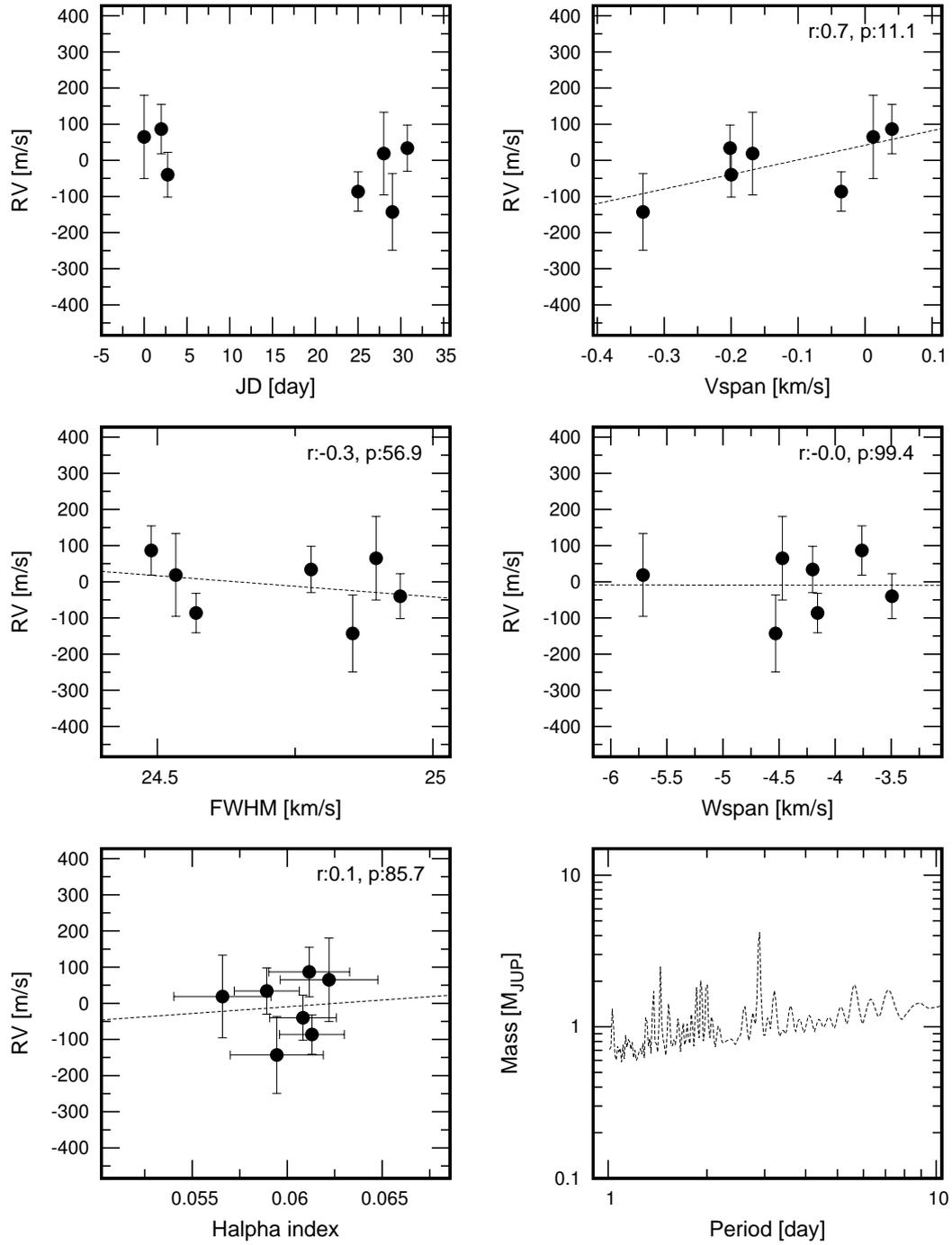}
\end{center}
\vspace{-0.7cm}
\caption{Same as figure \ref{appen1} for HIP 18091.}
\end{figure}

\newpage
\begin{figure}[htb]
\begin{center}
\includegraphics[width=1.0\textwidth]{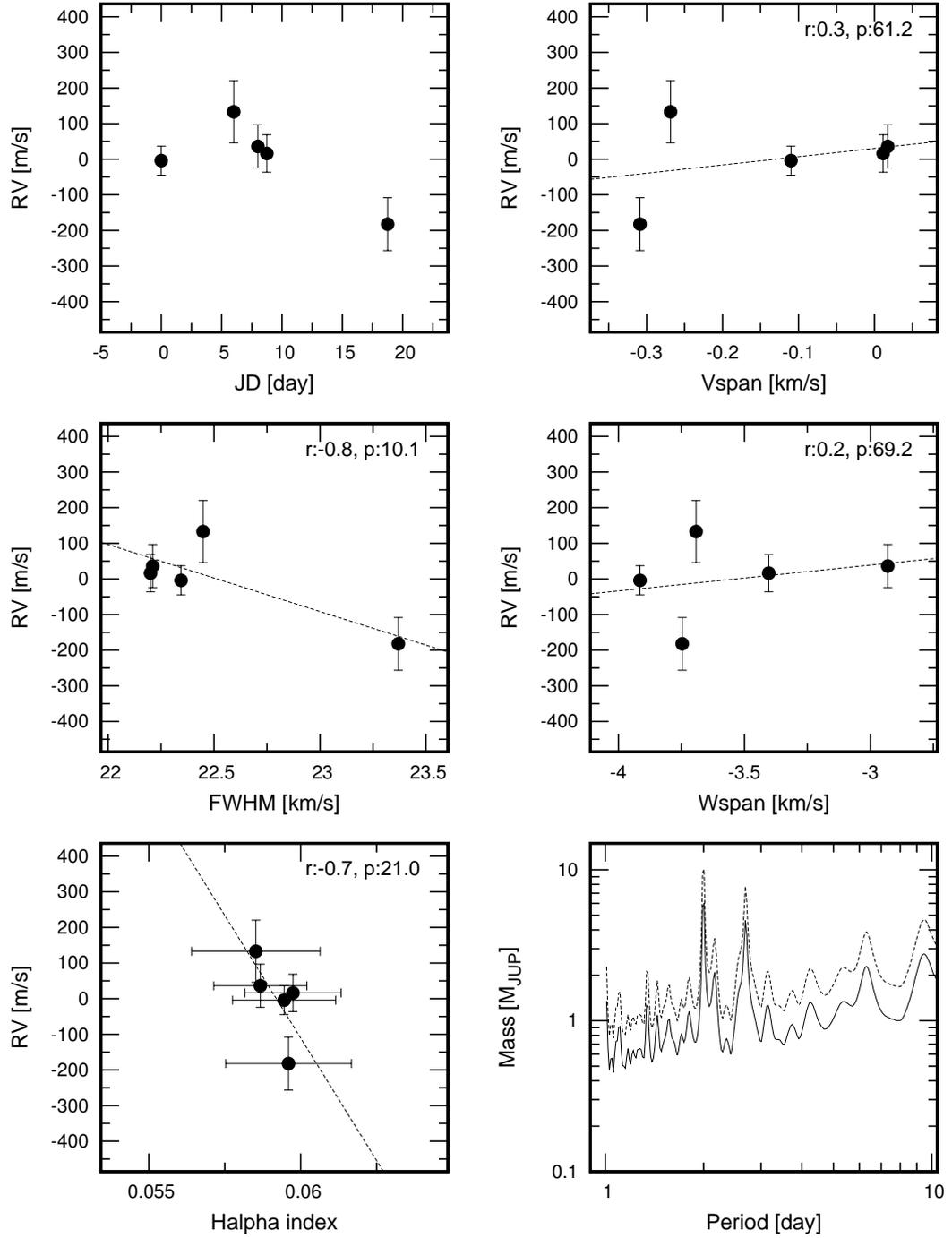}
\end{center}
\vspace{-0.7cm}
\caption{Same as figure \ref{appen1} for TYC 1247-515-1.}
\end{figure}

\newpage
\begin{figure}[htb]
\begin{center}
\includegraphics[width=1.0\textwidth]{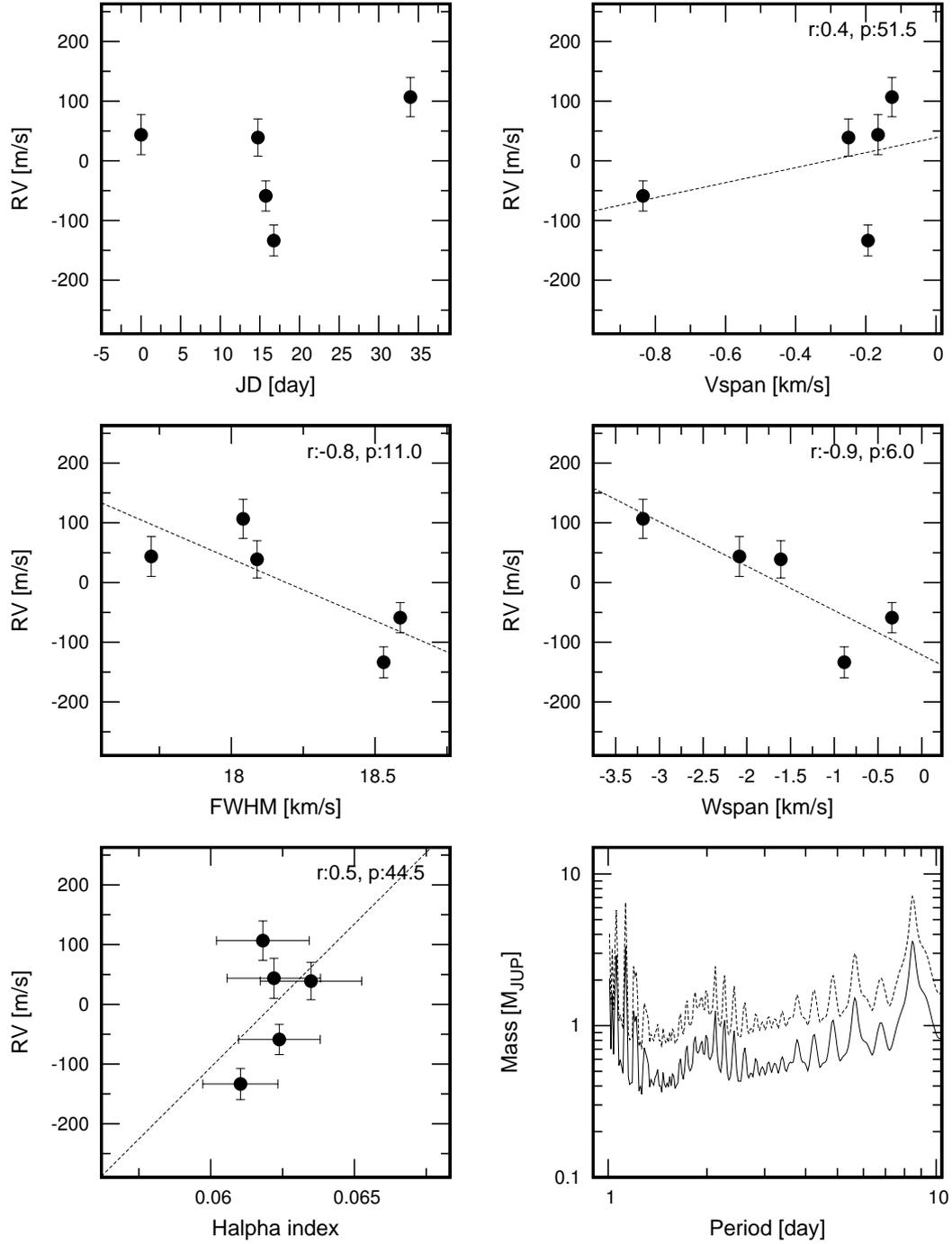}
\end{center}
\vspace{-0.7cm}
\caption{Same as figure \ref{appen1} for TYC 1260-1107-1.}
\end{figure}

\newpage
\begin{figure}[htb]
\begin{center}
\includegraphics[width=1.0\textwidth]{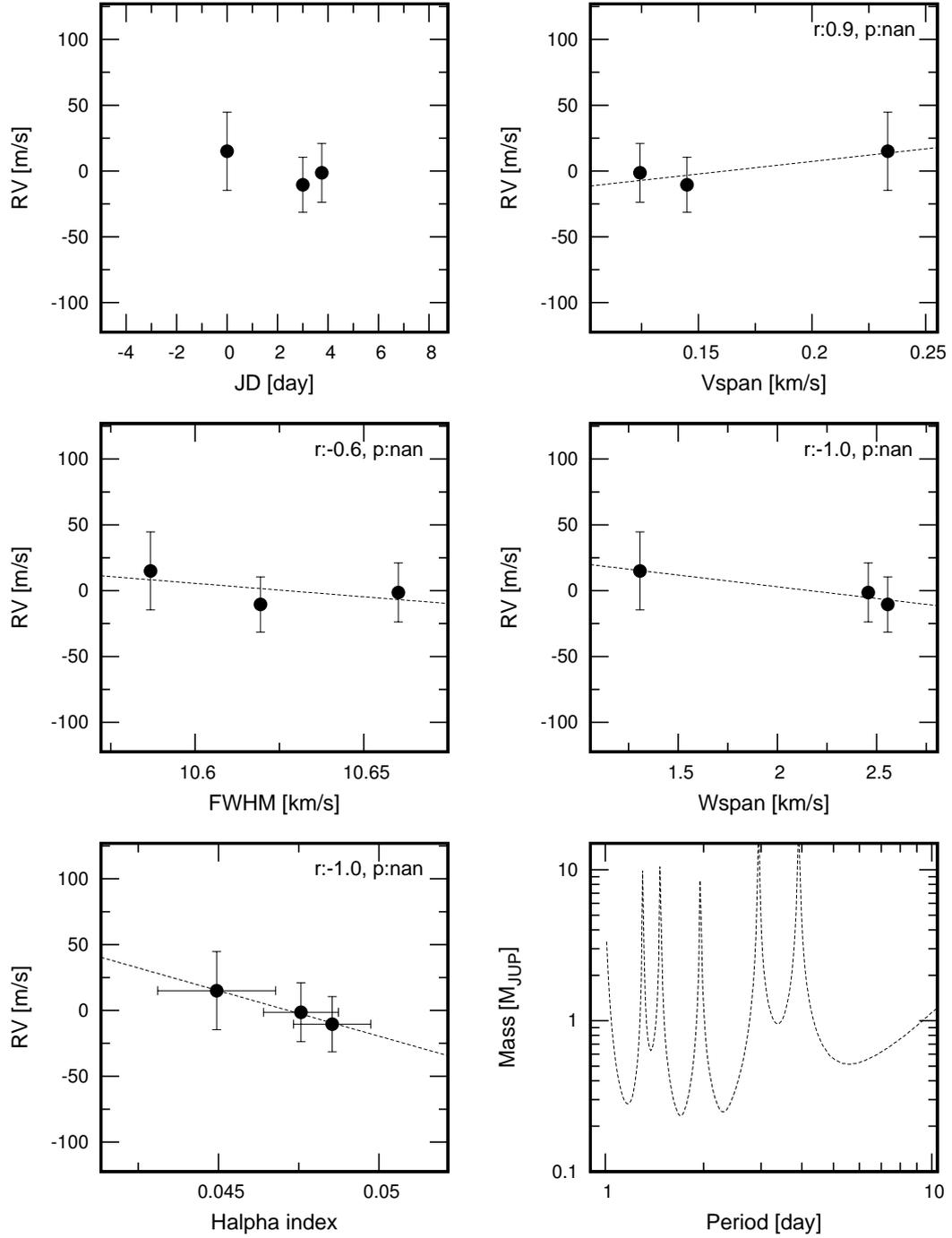}
\end{center}
\vspace{-0.7cm}
\caption{Same as figure \ref{appen1} for TYC 1797-1503-1.}
\end{figure}

\newpage
\begin{figure}[htb]
\begin{center}
\includegraphics[width=1.0\textwidth]{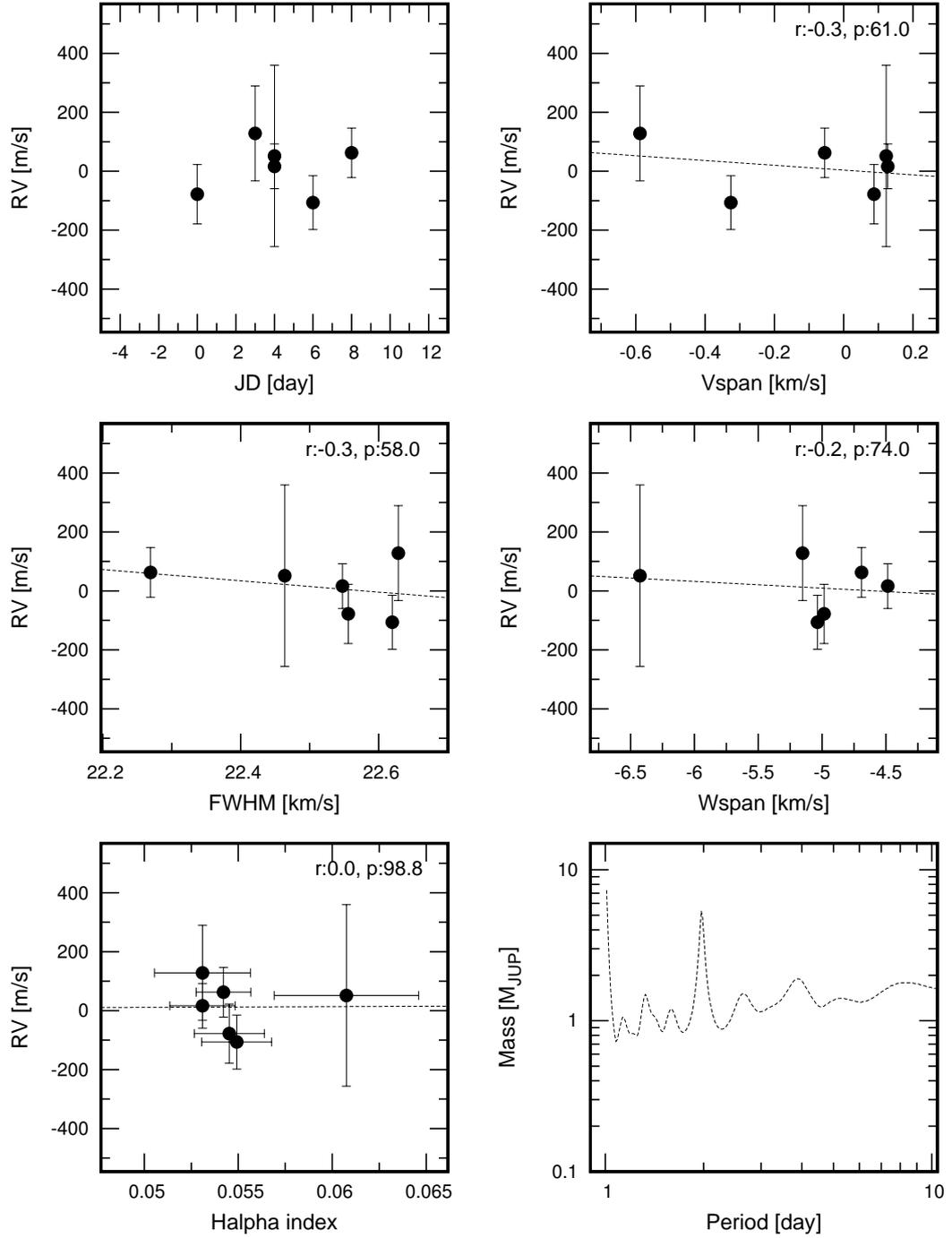}
\end{center}
\vspace{-0.7cm}
\caption{Same as figure \ref{appen1} for TYC 1798-465-1.}
\end{figure}

\newpage
\begin{figure}[htb]
\begin{center}
\includegraphics[width=1.0\textwidth]{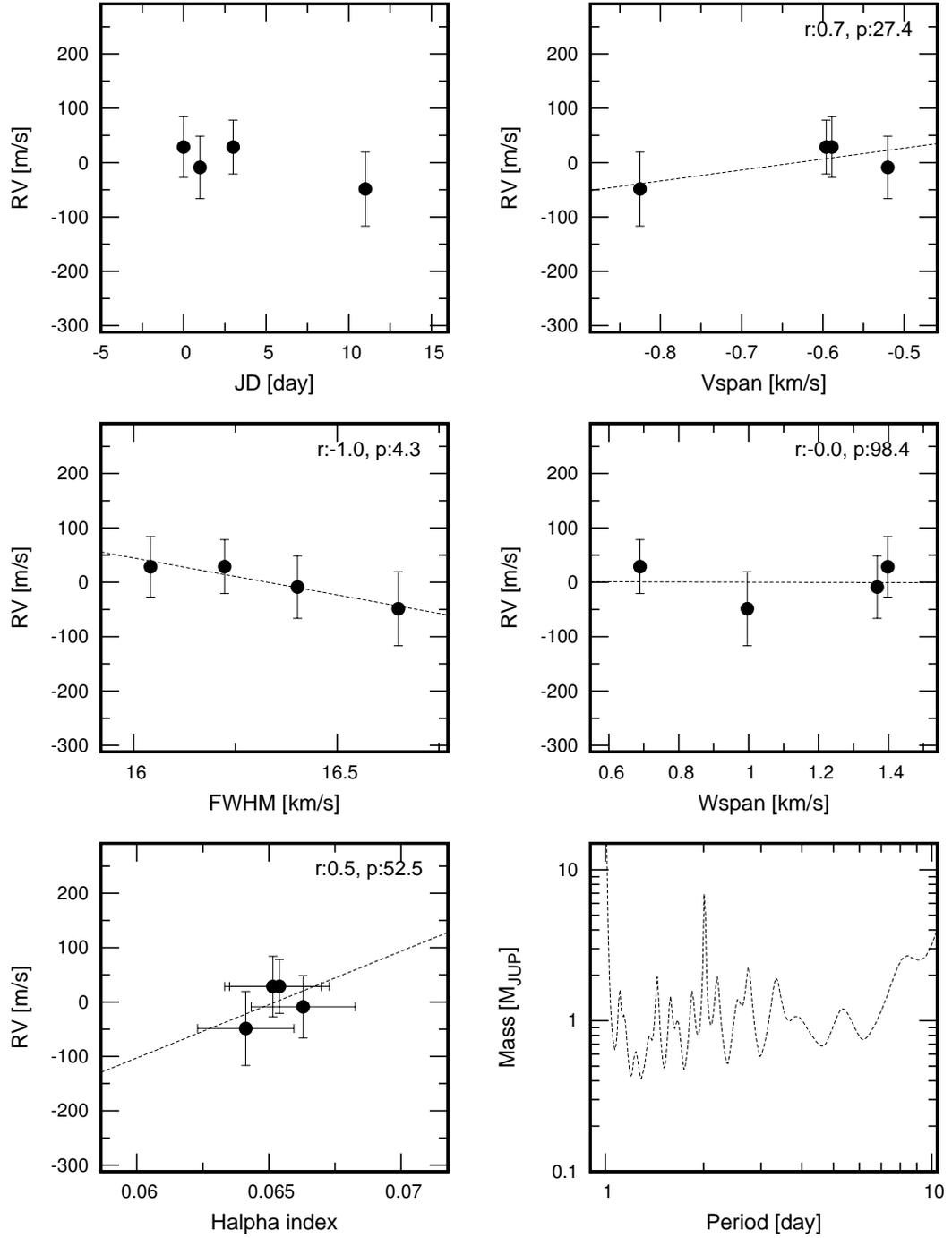}
\end{center}
\vspace{-0.7cm}
\caption{Same as figure \ref{appen1} for TYC 1799-757-1.}
\end{figure}

\newpage
\begin{figure}[htb]
\begin{center}
\includegraphics[width=1.0\textwidth]{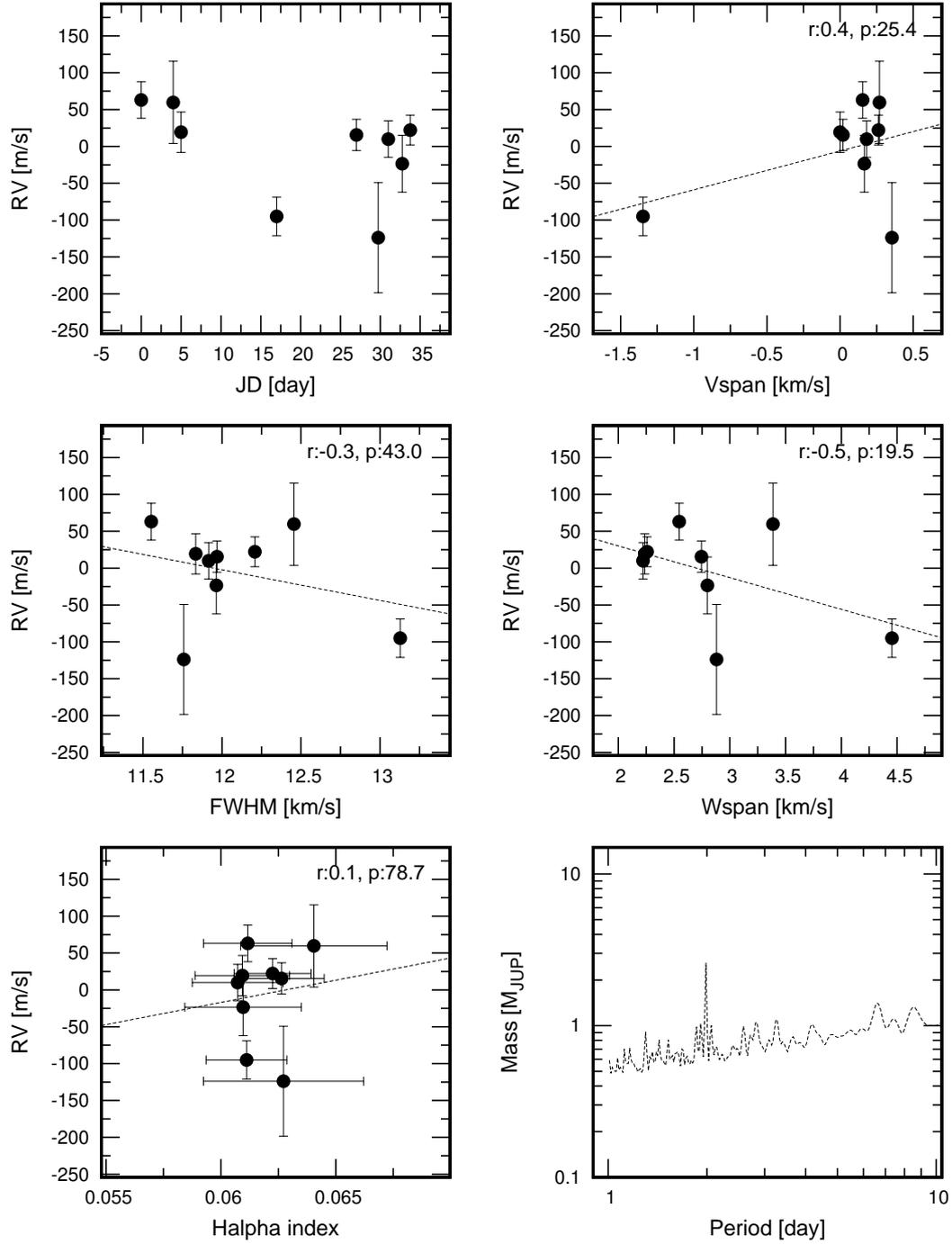}
\end{center}
\vspace{-0.7cm}
\caption{Same as figure \ref{appen1} for TYC 1800-1616-1.}
\end{figure}

\newpage
\begin{figure}[htb]
\begin{center}
\includegraphics[width=1.0\textwidth]{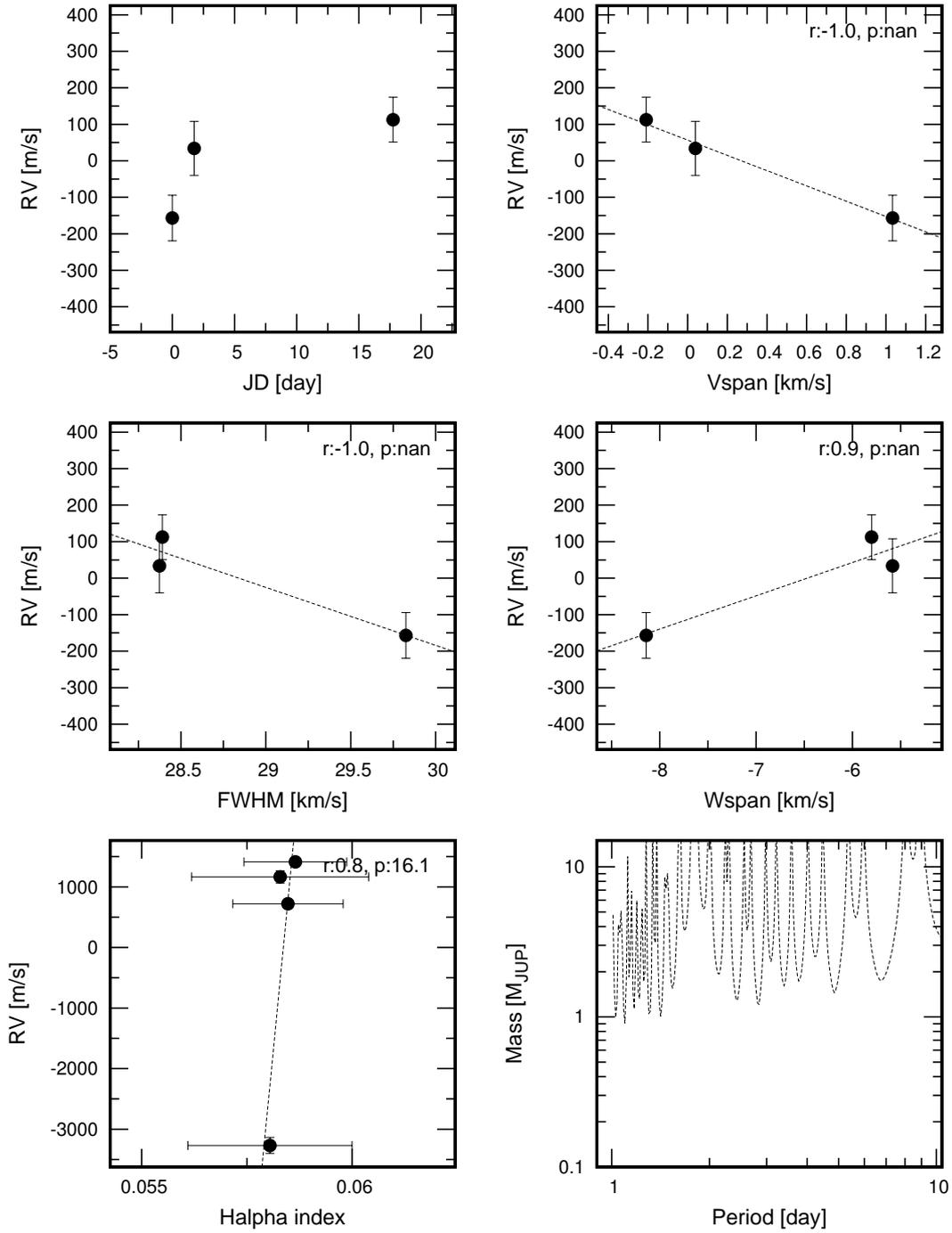}
\end{center}
\vspace{-0.7cm}
\caption{Same as figure \ref{appen1} for TYC 1800-1852-1.}
\end{figure}

\newpage
\begin{figure}[htb]
\begin{center}
\includegraphics[width=1.0\textwidth]{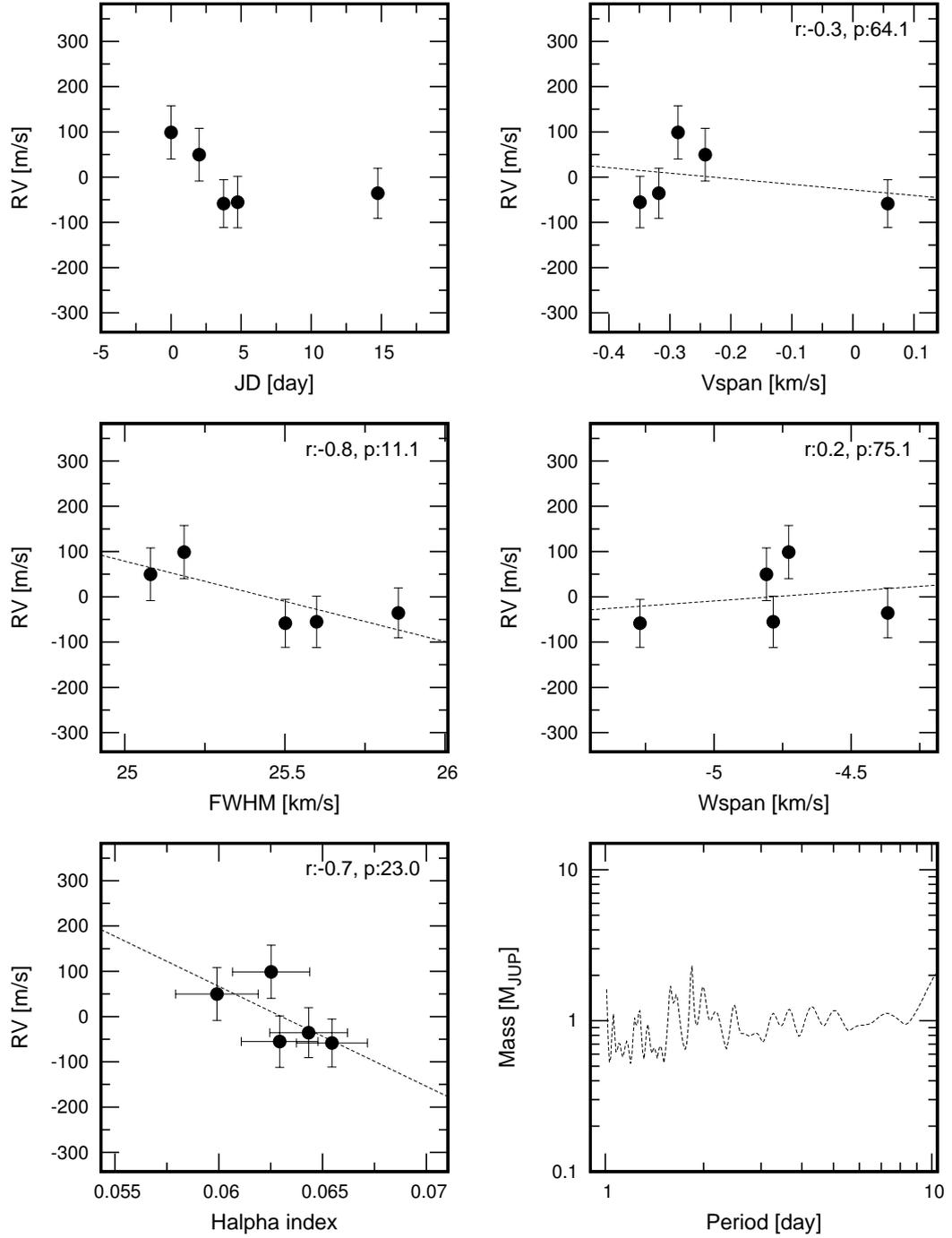}
\end{center}
\vspace{-0.7cm}
\caption{Same as figure \ref{appen1} for TYC 1800-1917-1.}
\end{figure}

\newpage
\begin{figure}[htb]
\begin{center}
\includegraphics[width=1.0\textwidth]{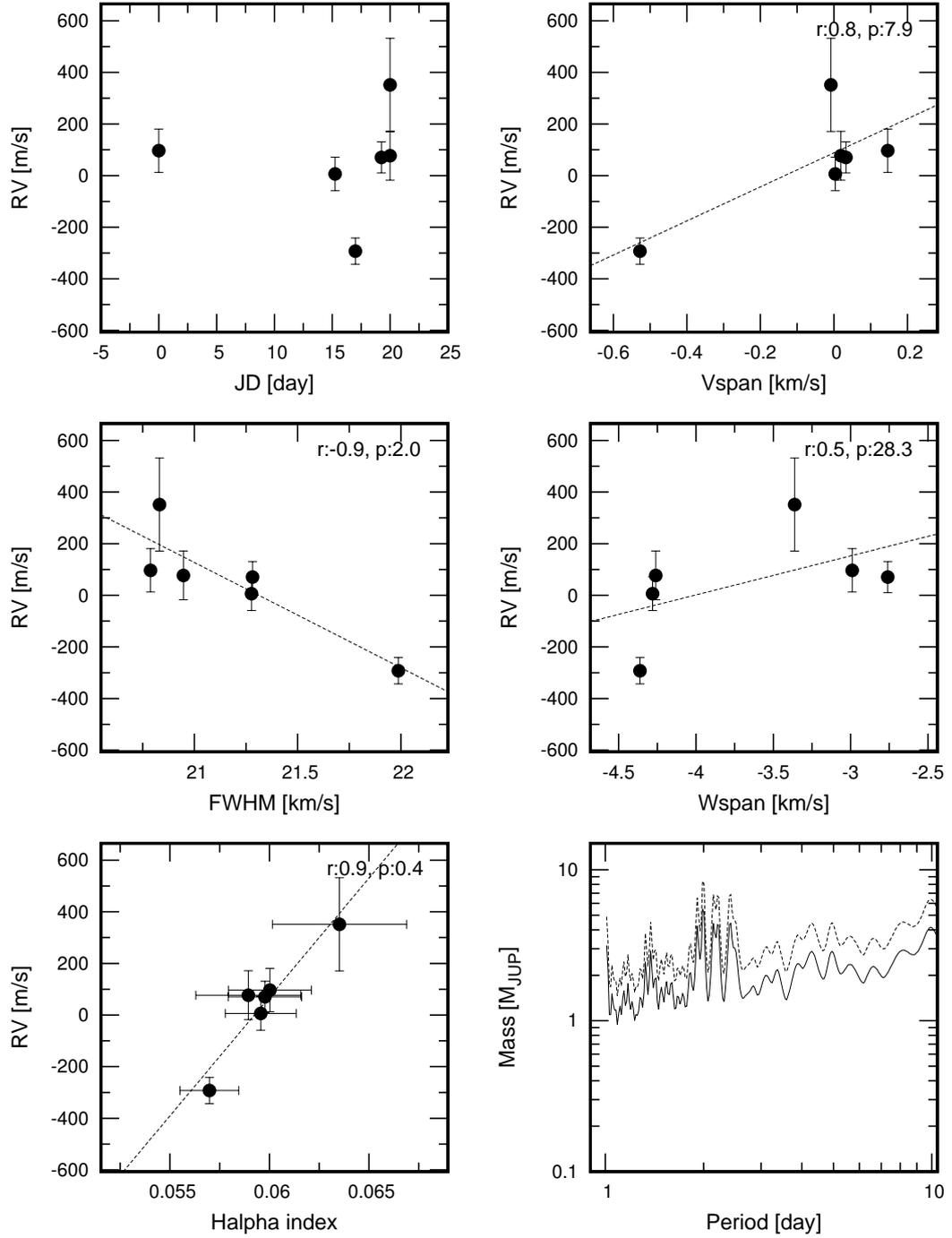}
\end{center}
\vspace{-0.7cm}
\caption{Same as figure \ref{appen1} for TYC 1800-471-1.}
\end{figure}

\newpage
\begin{figure}[htb]
\begin{center}
\includegraphics[width=1.0\textwidth]{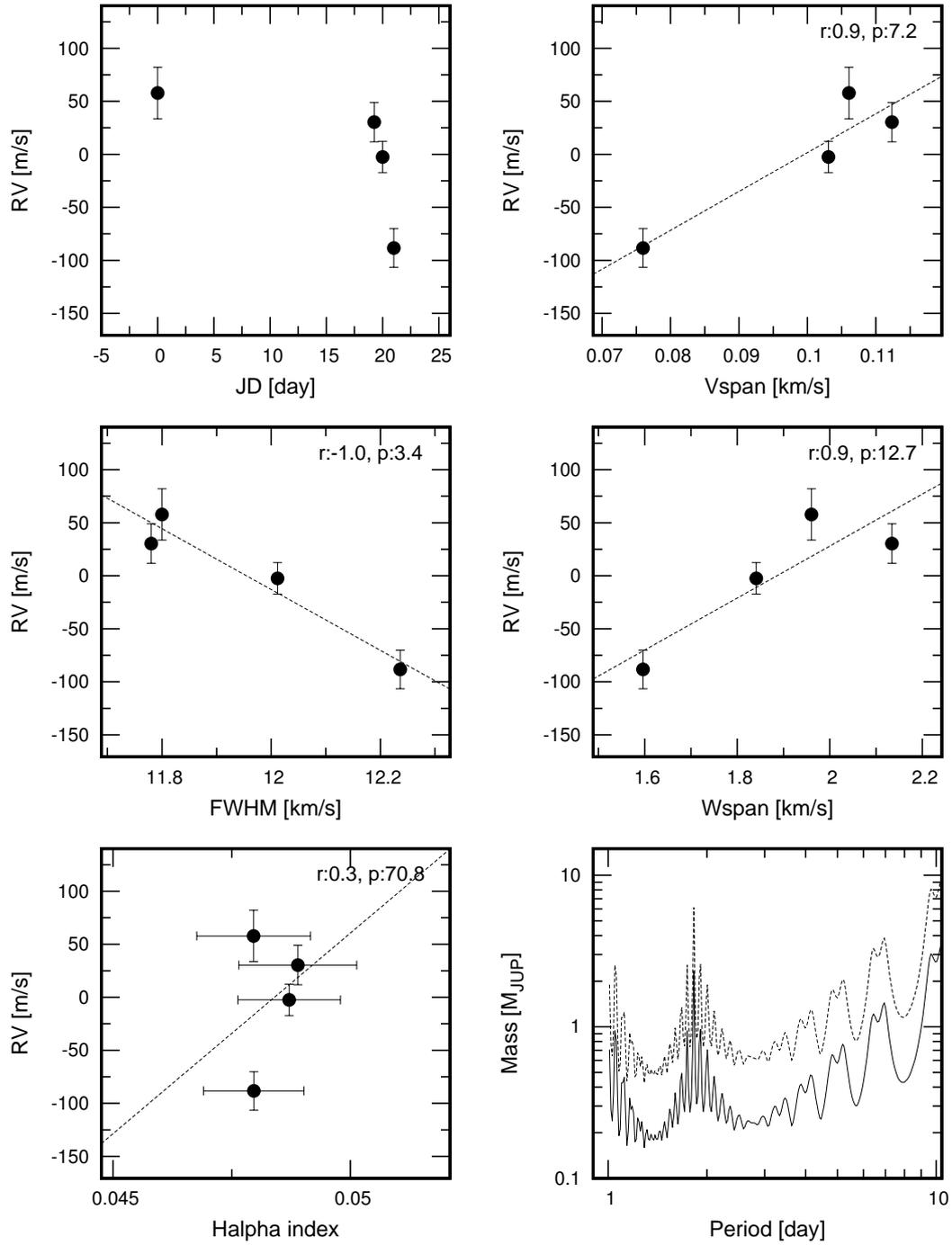}
\end{center}
\vspace{-0.7cm}
\caption{Same as figure \ref{appen1} for TYC 1803-697-1.}
\end{figure}



\end{document}